\documentclass[namedreferences,final]{SolarPhysics}
\usepackage[dvips]{graphicx}
\usepackage[optionalrh]{spr-sola-addons}
\usepackage{multirow}
\usepackage{rotating}
\usepackage{color}
\usepackage[T1]{fontenc}
\usepackage{times}
\usepackage{natbibmod}
\DeclareMathVersion{bold}
\usepackage{bm}
\usepackage[psamsfonts]{amsfonts}
\usepackage{mathptmx}
\usepackage{url}
\renewcommand{\thetable}{\arabic{table}}
\newcommand{\diff}[2]{\frac{\partial #1}{\partial #2}}
\renewcommand{\d}{\mathrm{d}}
\newcommand{\urlBiBTeX}[1]{\url{#1}}

\bibpunct[, ]{(}{)}{;}{a}{,}{,}

\begin{document}
\begin{article}
\begin{opening}

\title{Predicting Solar Flares by Data Assimilation in Avalanche Models.\\
I. Model Design and Validation}
\author{Eric \surname{Bélanger},
        Alain \surname{Vincent},
	Paul \surname{Charbonneau}}
\runningauthor{Bélanger {\it et al.}}
\runningtitle{Predicting Solar Flares by Data Assimilation}
\institute{Département de physique, Université de Montréal, \\
C.P. 6128 Succ. Centre-Ville, Montréal, Qc, H3C-3J7, CANADA.\\ 
  ({\it e-mails:} {\tt belanger@astro.umontreal.ca}, {\tt
    vincent@astro.umontreal.ca}, {\tt paulchar@astro.umontreal.ca})}
\date{Received ; accepted }

\begin{abstract}
  Data assimilation techniques, developed in the last two decades
  mainly for weather prediction, produce better forecasts by taking
  advantage of both theoretical/numerical models and real-time
  observations.  In this paper, we explore the possibility of applying
  the data-assimilation techniques known as 4D-VAR to the prediction
  of solar flares. We do so in the context of a continuous version of
  the classical cellular-automaton-based self-organized critical
  avalanche models of solar flares introduced by Lu and Hamilton
  ({\it Astrophys. J.}, {\bf 380}, L89, 1991). Such
  models, although {\it a~priori} far removed from the physics of magnetic
  reconnection and magneto\-hydrodynamical evolution of coronal
  structures, nonetheless reproduce quite well the observed
  statistical distribution of flare characteristics.  We report here
  on a large set of data assimilation runs on synthetic energy release
  time series. Our results indicate that, despite the unpredictable
  (and unobservable) stochastic nature of the driving/triggering
  mechanism within the avalanche model, 4D-VAR succeeds in producing
  optimal initial conditions that reproduce adequately the time series
  of energy released by avalanches/flares. This is an essential first
  step towards forecasting real flares.
\end{abstract}

\end{opening}

\section{Introduction: Flares and Self-Organized Criticality}

Spatially-resolved observations of solar flares have revealed a very
broad range of scales in the flaring phenomenon. 
Probability distributions of global flare characteristics such as peak
flux, energy release, and duration are now known to take the form of
power laws spanning many decades in size \citep[eight in the case of flare
energy; see, {\it e.g.},][]{dennis,lu93,aschwandenb}.  This is surprising
because the vast majority of flares occur in active regions and
activity complexes that have global characteristics (linear size,
magnetic flux, peak field strength) that are much more narrowly
distributed. This
indicates that the flaring phenomenon is intrinsically scale-free,
even though its energy reservoir may not be. The relatively-slow evolution
of active regions is also in stark contrast to the short energy
release timescale associated with the flaring phenomenon.

Avalanches are one class of physical phenomena that are characterized
by intermittent, scale-free energy release even under conditions of
slow, continuous energy loading. In the flare context, the physical
picture usually invoked is that of coronal magnetic structures being
slowly and stochastically
forced by photospheric fluid motions, leading to the gradual
buildup of electrical current sheets in the coronal
plasma~\citep{parker83a,parker88}. Plasma instabilities eventually
trigger magnetic reconnection at an unstable site, leading to
alterations of the physical conditions at neighbouring current sheets
that can then themselves become unstable, and so on in classical
avalanching style.  Numerous avalanche models inspired by this general
scenario have been developed to describe solar flares \citep[see][and references therein]{charbonneau}. Many of these models
manage to produce avalanche size distributions having the form of
power laws, with logarithmic exponent comparing fairly well to
observationally-determined values \citep[{\it e.g.},][]{lu93,georgoulisvlahos,aschwandenchar}.
Most of these models are based on the idea of self-organized
criticality (SOC)~\citep*{bak,jensen}.  The ``criticality'' is akin to
phase change in equilibrium thermodynamics, where the effects of a
small, localized perturbation can be felt on a dynamical timescale
throughout the
whole system. The system is said to be ``self-organized'' when this
critical state is a dynamical attractor and is reached in response to
external forcing without requiring fine tuning of a control parameter.
Generally, SOC is found in slowly-driven, open, dissipative systems
subjected to a self-limiting local threshold instability. The
threshold to the instability is crucial, as it allows the system to
transit from one metastable state to another while preventing the
dynamics to be governed by external forcing~\citep{jensen}. Potential
examples in the natural world include various forms of sandpiles,
avalanches and landslides, but also earthquakes, forest fires,
hydrological networks, traffic jams, magnetospheric substorms, and
solar flares \cite*[see][for a spirited exposition]{bak96}.

If flares are truly a manifestation of SOC dynamics, then the outlook
for accurate flare forecasting would appear, {\it a~priori}, pretty grim indeed.
Nothing fundamentally distinguishes a large flare from a small one,
flare size simply being a matter of the number of current sheets
involved in the avalanche of reconnection events. Even worse, the
triggers of flares large and small are the same, namely a small (quite
possibly unobservable) perturbation affecting the system somewhere
locally. However, the occurrence of a large avalanche is only possible
if a large, ``connected'' portion of the system is close to the
avalanching threshold. The state of the system, in turn, is a function
of its prior history, and in particular of the past occurrence of
avalanches, of which the larger ones are (presumably) observable. In
other words, past avalanching behaviour holds clues to the current
state of the system, and therefore to its {\it potential} avalanching
behaviour.

The question is then: can this information be retrieved and used to
produce reliable avalanche forecasts, despite the stochastic nature
of the driving/triggering mechanism? This is the central question we
address in this series of papers, using data-assimilation techniques.
This first paper describes the SOC avalanche model and 
data assimilation technique we are developing towards forecasting,
and demonstrates that the resulting scheme can adequately reproduce
the avalanching behaviour of the system even in the absence of detailed
information on the spatio-temporal behaviour of 
the stochastic driver. This is a first essential
step towards forecasting, which is the topic of the subsequent
papers in the series.

This first paper is organized as follows:
Section~\ref{sec:tube} gives an overview of a simple, ``classical'',
discrete SOC model based on a cellular automaton, as well as a
continuous analog described by a partial differential equation
reversed-engineered from the discrete cellular automaton rules. This
is the model used in Section~\ref{sec:4dvar} in conjunction with the
4D-VAR data assimilation techniques. Section~\ref{sec:result} presents
results for a wide set of validation experiments demonstrating that
4D-VAR can successfully reproduce the avalanching behaviour present in
energy release time series, despite the stochastic nature of the
external forcing that loads energy into the system and triggers
avalanches. 
We conclude in Section~\ref{sec:conclusion} by
summarizing the main results of this study, and outlining the road
lying ahead, towards forecasting real solar flares.

\section{Avalanche Models}
\label{sec:tube}

\subsection{The Lu and Hamilton Model}
\label{sec:soc}

The sandpile has now become the icon of SOC systems~\citep*{bak}.
As sand grains are dropped one by one on a flat surface, a sandpile
will build up, with occasional avalanches of various sizes, until
the pile has reached a conical shape with the slope everywhere at 
or near the angle of repose. Addition of more sand grains can now trigger
large avalanches disrupting the whole slope, or the toppling of only
a few sand grains, or nothing at all. The system has reached a
statistically stationary state where, averaged over a long-enough temporal
interval, as many sand grains fall off the pile as are dropped on it.
Notice that while the loading is slow and gradual, the unloading
is strongly intermittent and involves avalanches of all sizes.

The statistical physics of sandpiles has been extensively studied
using cellular automata models, where the sandpile is replaced by a
lattice of locally interconnected nodes on which a nodal variable
related to energy is defined~\citep{kadanoff}.  In the context of solar
flares, the first such model is to be found in the groundbreaking work
of~\cite{lu91} and \cite{lu93} \cite[but see also][]{zirker}. Consider the
following two-dimensional scalar version of the~\cite{lu91} model; a
scalar nodal quantity ($A_{i,j}^n$) is defined over a $N\times N$
regular cartesian grid with nearest-neighbour connectivity
\cite[top+down+right+left; see, {\it e.g.}, Figure~1 in][]{charbonneau}.  Here
the superscript $n$ is a discrete time index, and the subscript pair
($i,j$) identifies a node on the 2D lattice. The cellular automaton is
driven by adding small increments in $A$ at randomly selected nodes in
the lattice (one per time step), analogous here to dropping sand
grains on the pile.  A stability criterion is defined in terms of the
local curvature of the field at node $(i,j)$:
\begin{equation}
  \label{eq:stab1}
  \Delta A_{i,j}^n\equiv A^n_{i,j}
  -\frac{1}{4}\sum_{\mathrm{neighbours}}A_{\mathrm{neighbours}}^n~,
\end{equation}
where the sum runs over the four nearest neighbours at
nodes \mbox{$(i,j\pm 1)$} and \mbox{$(i\pm 1,j)$}.
If this quantity exceeds some pre-set threshold ($A_c$) (analogous
to the slope exceeding the angle of repose on the sandpile), then
$A_{i,j}^n$ is redistributed to its four nearest neighbours
according to the following rules:
\begin{equation}
  \label{eq:redis1}
  A_{i,j}^{n+1}= A_{i,j}^n-\frac{4}{5}\Delta A_{i,j}^n~,
\end{equation}
\begin{equation}
  \label{eq:redis2}
  A_{i\pm 1,j\pm 1}^{n+1}= A_{i\pm 1,j\pm 1}^n+\frac{1}{5}\Delta A_{i,j}^n~.
\end{equation}
These redistribution rules are conservative in $A$; however,
if one identifies $A^2$ with a measure of energy (more on this below),
then it is readily verified that they lead
to a decrease of the total lattice energy:
\begin{equation}
  \label{eq:elatt}
  E_l^n=\sum_{i,j}({A_{i,j}^n})^2~,
\end{equation}
the excess energy being what is liberated in the ``flare''.
\cite{lu91} and \cite{lu93} have shown that this driven cellular automaton can
produce avalanches with robust power-law exponents resembling those
inferred from flare observations \cite[see also][and
references therein]{charbonneau}.

Giving a physical meaning to the scalar $A$ is not trivial.  If $A$ is
considered as being the magnetic field, then $\nabla\cdot \mathbf{A}\neq 0$.
Taking $A$ as the vector potential automatically solves the non-null divergence
problem.  However, $\sum_{i,j}({A_{i,j}^n})^2$ is then no longer an obvious
measure of magnetic energy~\citep{charbonneau}.  Since
$\nabla\times\mathbf{A}=\mathbf{B}$, dimensional analysis does suggest 
$|\mathbf{A}|^2\sim L^2|\mathbf{B}|^2$ so we can at first assume that the vector
potential scales with the magnetic field, a conjecture supported by
the numerical results of \cite*{isliker}, using an avalanche model
closely related to that described above. Variations in lattice energy ($E_l$)
from one time step to the next during an avalanche then yields the energy
liberated at each time step:
\begin{equation}
  \label{eq:erel}
  \begin{array}{ll}E_r^n\end{array}=\left\{\begin{array}{ll}E_l^{n-1}-E_l^{n}~(>0),&\qquad\mathrm{lattice~avalanching}\\
      0,&\qquad\mathrm{otherwise.}\end{array}\right.
\end{equation}
The resulting time series of energy release
is the target for data assimilation. In the Lu and Hamilton discrete
model, $E_r$ can be calculated analytically knowing the redistribution
rule and stability measure, but this will no longer be the case
with the continuous analog to be introduced shortly; Equation~(\ref{eq:erel})
applies equally well to both classes of models.

At this juncture it must be emphasized that the stochasticity in the
Lu and Hamilton
model (and other published variations thereof)
is not a mere noise-like ``inconvenience'' superimposing itself
on an underlying deterministic flaring process. In the Lu and Hamilton
SOC avalanche model, stochastic forcing plays the dual role of energy 
loading {\it and} avalanche triggering. The model thus has a fundamentally
stochastic component.

\subsection{A Continuous SOC Model}

\subsubsection{Reverse Engineering}

The structure of Equations~(\ref{eq:redis1}) and
(\ref{eq:redis2}) suggests that they can be interpreted as 
the result of applying centered, second-order, finite difference
and a one-step, explicit, time-stepping algorithm to some
partial differential equation (PDE) in two spatial dimension
discretized on a regular cartesian grid. Applying this
reverse engineering approach \cite[see also][]{liu}
leads to the following PDE describing the evolution of $A$
during an avalanche:
\begin{equation}
  \label{eq:ava}
  \diff{A}{t}=-\diff{^2}{x^2}\left(\nu(\nabla^2A)\diff{^2 A}{x^2}\right)
  -\diff{^2}{y^2}\left(\nu(\nabla^2A)\diff{^2 A}{y^2}\right)~.
\end{equation}
Likewise, the RHS of Equation~(\ref{eq:stab1}) is readily interpreted
as a second-order centered finite-difference representation of
the Laplacian operator acting on $A$, so that the diffusion
coefficient $\nu(\nabla^2A)$ appearing in Equation~(\ref{eq:ava}) is given by:
\begin{equation}
  \label{eq:ava1}
  \begin{array}{ll}\nu(\nabla^2A)\end{array}
  =\left\{\begin{array}{ll}\nu_a, &\mathrm{if}\ (\nabla^2A)^2>A_c^2\\
  0, & \mathrm{otherwise}\end{array}\right.
\end{equation}
where $A_c$ is the stability threshold.  The numerical value of
$\nu_a$ depends on the redistribution rules that were used in the
discrete equations (Equations~(\ref{eq:redis1}) and (\ref{eq:redis2})) and
on the grid size. In the case of the Lu and Hamilton model, in two
spatial dimensions,
$\nu_a=\Delta^2/20$ ($\Delta x=\Delta y\equiv\Delta$), with units of
[time]/[length]$^2$ implicitly included in the denominator,  a value used
in all calculations reported
upon below.  In analogy with the discrete model, we identify the
functional $A(x,y,t)$ with a measure of the magnetic vector potential,
and assume that $A^2$ is a measure of energy.
 
Equation~(\ref{eq:ava}) is a fourth-order ``hyper-diffusion''
equation, albeit a strongly nonlinear one since the hyper-diffusion
coefficient is only non-zero when the system is avalanching, and even
then it remains a discontinuous function of position, being only
non-zero at unstable nodes. If all nodes are stable, then the quantity
$A$ evolves in response to the external forcing only.  In the
classical cellular automaton, at each (non-avalanching) time step, a
small increment of random amplitude in $A$ is added to a single
randomly selected node of the lattice. Reverse-engineering of this
forcing rule immediately leads to:
\begin{equation}
  \label{eq:ava2}
  \diff{A}{t}= F_R(x_0(t),y_0(t))
\end{equation}
where the forcing location $(x_0,y_0)$ varies randomly in time. The
continuous model then amounts to solving Equation~(\ref{eq:ava2}) when the
system is everywhere stable, and switching to Equation~(\ref{eq:ava}) if
one or more nodes become unstable. Note that this implies that the
forcing ($F_R$) is turned off during avalanches. This is the same
procedure used in the cellular automaton, and amounts to assuming that
there exists a wide separation of timescales between forcing and
avalanching, which in fact is well justified in the solar coronal
context \cite[see, {\it e.g.},][]{lu93}.

\subsubsection{Sample Numerical Results}

We now proceed to solve numerically the
continuous-avalanche equation (Equations~(\ref{eq:ava})\,--\,(\ref{eq:ava2})), by discretizing 
it using a centered finite-difference scheme with forward Euler
differencing for the time derivative. The reader may rightfully
wonder why then did we bother reverse-engineering the discrete Lu and Hamilton
model in the first place, but the need to proceed in this way
will become clear presently. Moreover, in solving
Equation~(\ref{eq:ava}) numerically as one would any partial
differential equations, subtle differences are introduced
with respect to the truly discrete model, and these 
must be clarified.

A square domain of linear size $L=2\pi$ is partitioned
using a regular cartesian grid ($\Delta x=\Delta y\equiv\Delta$).
Imposing $A=0$ on domain boundaries,
we solve dimensionless forms of the avalanche equations (Equations~(\ref{eq:ava})\,--\,(\ref{eq:ava2})), using
the hyper-diffusion time scale
\begin{equation}
  \label{eq:tscale}
  \tau_\nu=\frac{L^4}{\nu_a}
\end{equation}
as a time unit. This corresponds to the time it would take an avalanche
to sweep across the length of the domain.  Following a dimensional
analysis of the continuous avalanche equation, 
 $x$ and $y$ will now be in $L$ units and $t$ in $\tau_\nu$ units.  The diffusion
coefficient $\nu$, in the non-dimensional equation, is equal to 0 (stable)
or to 1 (avalanching).
For the scheme to remain numerically stable, a
von~Neumann stability analysis~\citep{press} indicates that the
dimensional time step must be chosen such that
\begin{equation}
  \label{eq:vonN}
\Delta t \le\frac{\Delta^4}{4\nu_a}
\end{equation}
or, in terms of dimensionless time:
\begin{equation}
  \label{eq:vonN2}
\frac{\Delta t}{\tau_\nu} \le\frac{1}{4}\left({\Delta\over L}\right)^4~.
\end{equation}
For our working $48\times 48$ mesh, $\frac{1}{4}\left({\Delta\over
    L}\right)^4=4.7\times 10^{-8}$, so a non-dimensional
time step of size $5.5\times 10^{-11}$
safely satisfies that condition, and is used for all
simulations unless otherwise specified.

The stability threshold is set at $A_c=7.0$.
Sequences of uniform random
deviates are used to define the stochastic forcing.
Both the location and magnitude of the
perturbation are randomly determined, the former excluding
boundary nodes and the latter constrained
to the pre-set interval \mbox{$[-1\times 10^{-5}:4\times 10^{-5}]$}. The perturbations
have non-zero mean to ensure buildup of $A(x,y)$ from the initial
state $A(x,y)=0$ throughout. This is
again in direct analogy to the Lu and Hamilton model described earlier.
Because the continuous model is discretized with centered second-order
finite differences, boundary conditions for the two rows of grid
points along the boundary are needed. The grid points along the
boundary are set to $A=0$ (as in the discrete model) while the
second row is set to $\nabla^2A=0$ through the use of the fictional-point
method~\citep*{mitchell}.  

\begin{figure}
  \centering 
\includegraphics[scale=0.9]{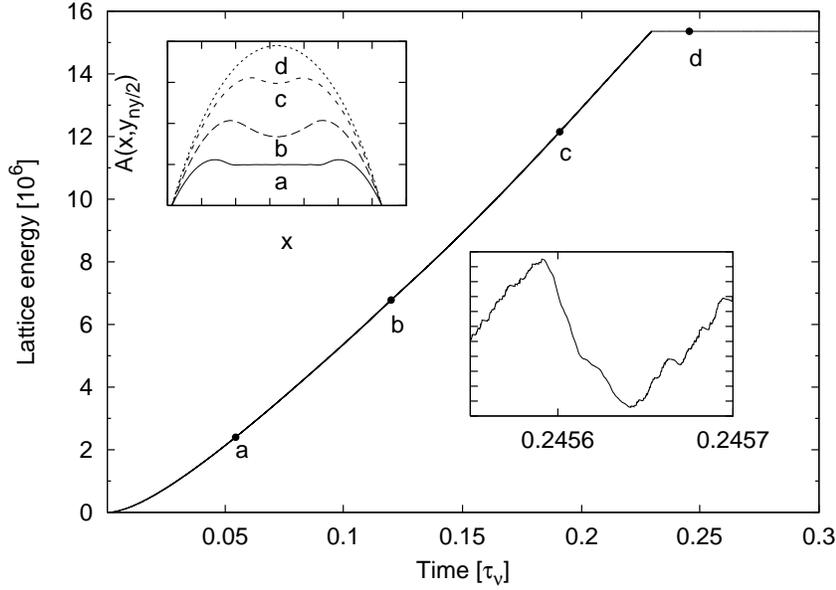}
\caption{Evolution of the lattice energy as a function
  of time. The lattice energy increases from the null state ($A=0$) to
  the SOC state as the energy input by the perturbations is greater
  than the energy released in avalanches.  At $t\simeq0.23$, the
  SOC state is reached: the energy added to the system is released by
  avalanches, hence the plateau.  The top left inset is a
  cross-section of the lattice variable $A$ along the x-axis, taken at
  different times.  The correspondingly labeled solid dots along the lattice energy curve indicate
  when the cross-sections were taken.  The bottom right inset is a
  zoom of the lattice energy curve about point ``d'', and shows the
  energy variation in the SOC state, with vertical axis covering five energy units.}
\label{fig:soc1}
\end{figure}

\begin{figure}
  \centering 
\includegraphics[scale=0.8]{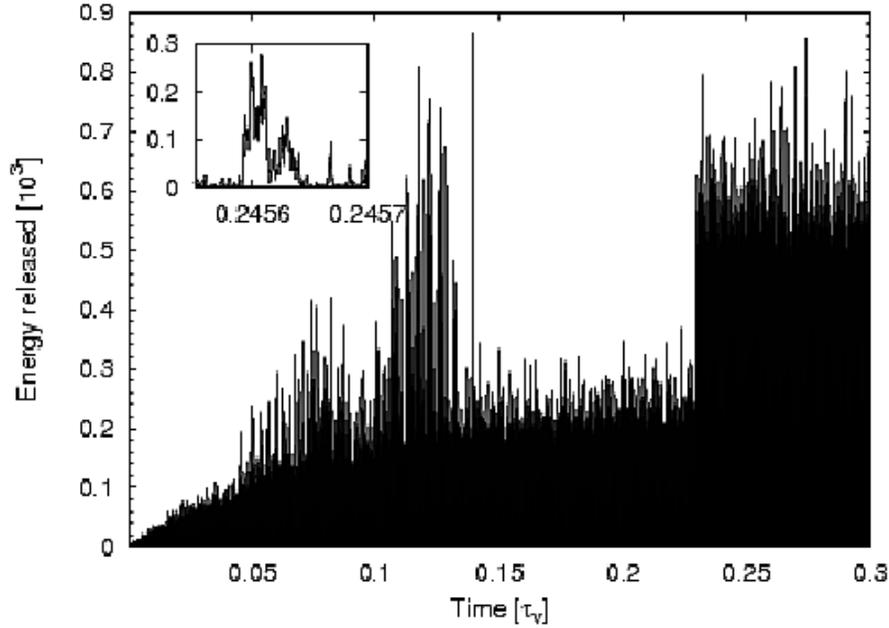}
\caption{Evolution of the energy released as a function of time.
  The large avalanches between $t\simeq0.1$ and
  $t\simeq0.14$ are due to a transient SOC state that is created
  when the stability criterion of the central region takes a value of
  $\sim +A_c$ while the periphery have $\sim -A_c$. This is only a
  transitional step as the center will fill up to take its final shape
  of an inverse parabola (``b''$\rightarrow$``d'' on the inset in
  Figure~\ref{fig:soc1}).  The inset is a zoom to show that we have
  individual avalanches separated by calm periods.}
  \label{fig:soc2}
\end{figure}

The first task is to bring the system to the SOC state.  Starting with
$A=0$ everywhere, the system is driven to the SOC state by the
external forcing, interrupted by avalanching episodes whenever and
wherever the stability threshold is exceeded.  Figure~\ref{fig:soc1}
illustrates the gradual buildup of lattice energy ($\sum A^2$) with
time. Initially, the energy gained by the system from the
perturbations exceeds the energy released in avalanches.  Thus, there
is a net increase of the system's energy.  However, at $t\simeq0.23$ a plateau begins.  The system has now reached a
statistically-stationary state where, in a time-averaged sense, the
buildup of $A$ in response to forcing is balanced by $A$ evacuated via
the $A=0$ boundary conditions when avalanches reach the system's
boundaries. Because large avalanches are more likely to reach the
boundaries, their frequency of occurrence jumps markedly once the
stationary state is reached ({\it cf.} Figure~\ref{fig:soc2}). This stationary
state is the SOC state.  The bottom 
right inset on Figure~\ref{fig:soc1} shows a zooming on the behaviour of the lattice
energy during the SOC state in the vicinity of point ``d''.  Every
decrease is the signature of an 
avalanche and the pronounced drop starting at $t=0.2456$ corresponds to
the large avalanche in the inset of Figure~\ref{fig:soc2}. The top left
inset shows cross-sections along the $y$-axis of 
the evolution of the scalar variable $A$ at different times up to the
SOC regime where it approximates the shape of an inverse parabola.  A
cross-section along the $x$-axis would yield the same result. All DNS
and 4D-VAR runs discussed further below are carried out with the
system in the SOC state.

Figure~\ref{fig:soc2} displays the energy released as a function of
time.  It is only in the
SOC state that avalanches spanning the whole system can be produced
with a significant frequency. For this reason, it will
be particularly important to ensure that the system is kept in the
SOC state throughout the data assimilation process. The large avalanches
between $t\simeq0.1$ and $t\simeq0.14$ (Figure~\ref{fig:soc2})
are due to a transient SOC state that is created when the stability
criterion of the central region takes a value of $\sim +A_c$ while the
periphery have $\sim -A_c$ (such as for the curve ``b'' on the top
left inset
of Figure~\ref{fig:soc1}). This is only a transitional step as the
center will fill up to eventually take its final shape of an inverse
parabola and the system will then be in its true SOC state.

Even though the avalanche model is a continuous one, it retains some
discrete features. Usually, in the numerical solution of
partial differential equations, reducing the time steps by a factor of
two but using twice as many of these smaller time steps, should not
affect the outcome significantly.  However, this is
not the case here, because of 
the spatio-temporally discrete nature of the
hyper-diffusion coefficient, as the latter is only turned on when the system
is unstable.  When this happens, the run with the smaller time step ($\Delta t$)
will redistribute more finely, as seen in Equation~(\ref{eq:ava}). When
stability is recovered, the formerly avalanching portion of the lattice
will find itself closer to the stability
threshold, thus the lattice energy will increase significantly.  The redistribution
is also affected by the value of the diffusion coefficient (with a
dependency on the distance between the grid points) which must be lower
for higher-resolution grids otherwise too much energy is redistributed
and the system restabilizes too far below the stability threshold, thus
leading to loss of SOC at the expense of large, quasi-periodic avalanches.

\begin{figure}
  \centering
  \includegraphics[scale=0.6]{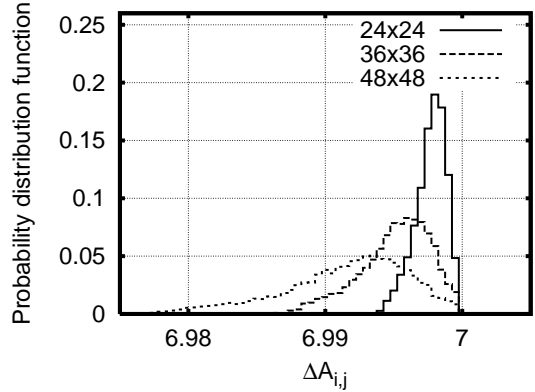}
  \caption{Normalized frequency distributions for the stability
    criterion evaluated at each grid point for runs of $24\times 24$, $36\times 36$,
    and $48\times 48$.  Data taken at ten different times, when the system
    was stable, were combined to generate the PDF. As expected, the
    peaks of the PDF lie close but below the stability threshold of
    $A_c=7.0$. As the resolution increases, the 
    distribution broadens and moves further from the threshold.  This is
    due to the increase of fine structures which permits more degrees
    of freedom to the large-resolution systems, thus making it easier
    to stray away from the threshold.}
  \label{fig:pdfda}
\end{figure}

The last difference between the Lu and Hamilton and continuous models pertains to
the effect of an increase in the number of grid points.  In the Lu and Hamilton
model, the distance between each grid point is constant ($\Delta=1$)
which leads to an increase of the size of the domain when the
grid point number is increased.  As for the continuous model, even
though the domain size can be increased, we will keep it constant in
this paper so an increase in the number of grid points will increase
the resolution.  The system will then be more effective in reproducing
fine structures as seen in Figure~\ref{fig:pdfda}, which shows the
normalized frequency distributions for the stability criterion
(Equation~(\ref{eq:stab1}))
evaluated at each grid point at ten different times for grid
resolution  of $24\times 24$,
$36\times 36$, and $48\times 48$. As with the discrete model, at any given time only
a small fraction of nodes are very close to the stability threshold.
The distribution is broader and centered further below
the stability threshold for the higher resolution runs. The reason
behind this behaviour is that the larger number of degrees of freedom
in the high-resolution run allows for the presence of finer structures,
which make it easier for the lattice to exceed the stability threshold, in the sense
that fluctuations of $\Delta A_{i,j}^n$ about its lattice mean
become larger as the spatial resolution is increased.

\subsubsection{Characteristics of Avalanches in the SOC State}

\begin{figure}
  \centering
  \includegraphics[scale=0.55]{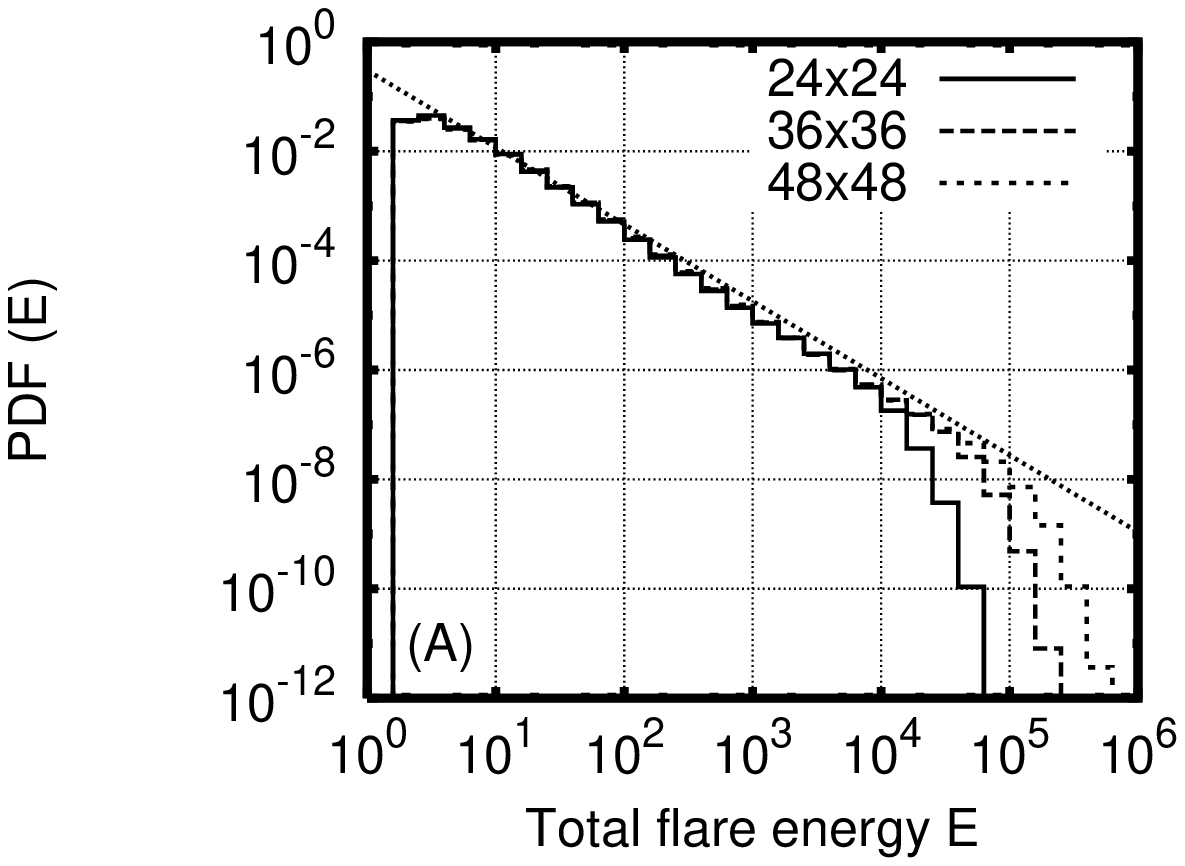}
  \includegraphics[scale=0.55]{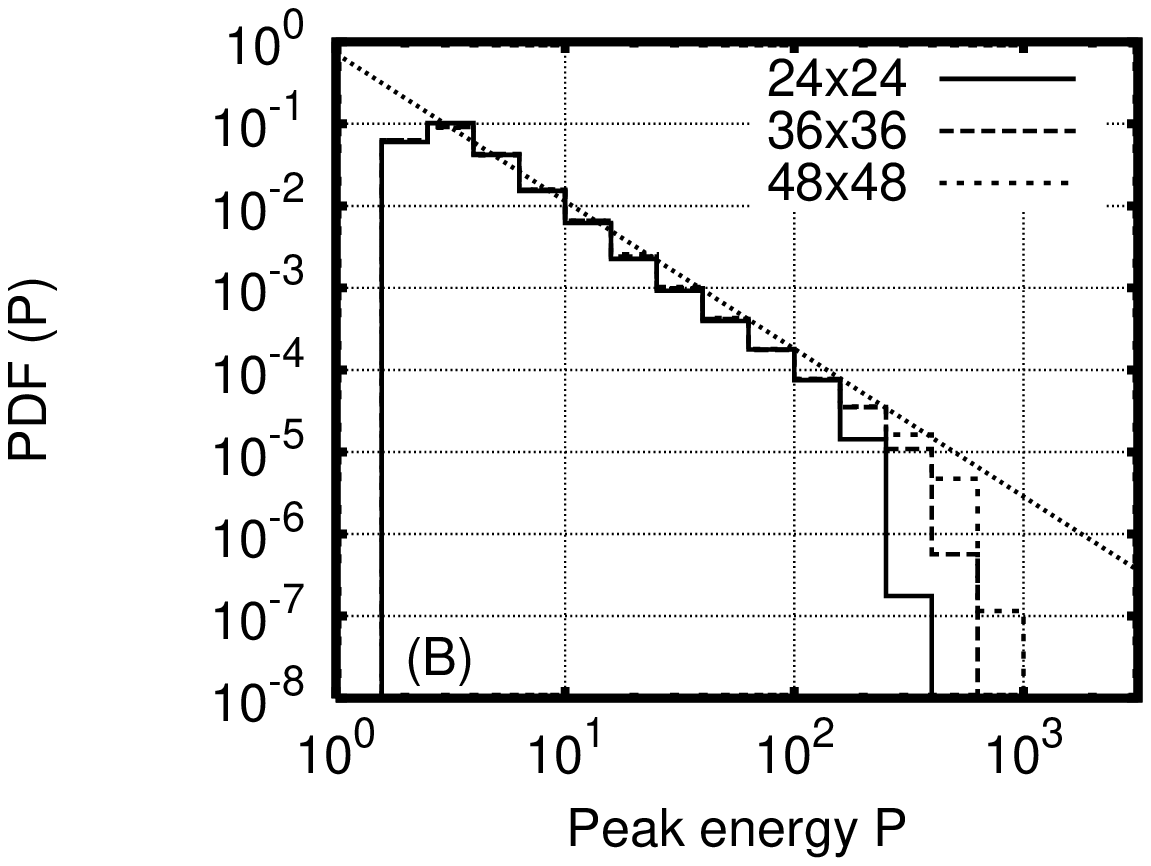}
  \includegraphics[scale=0.55]{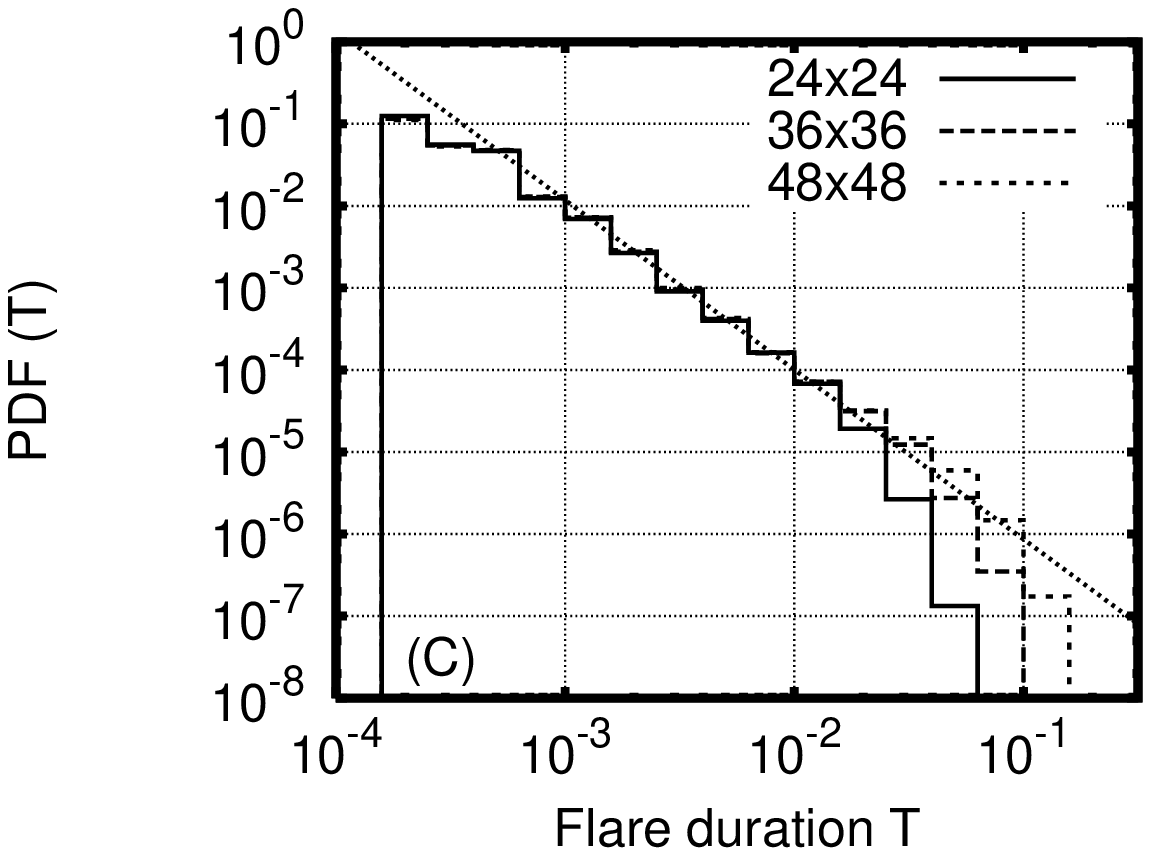}
  \includegraphics[scale=0.55]{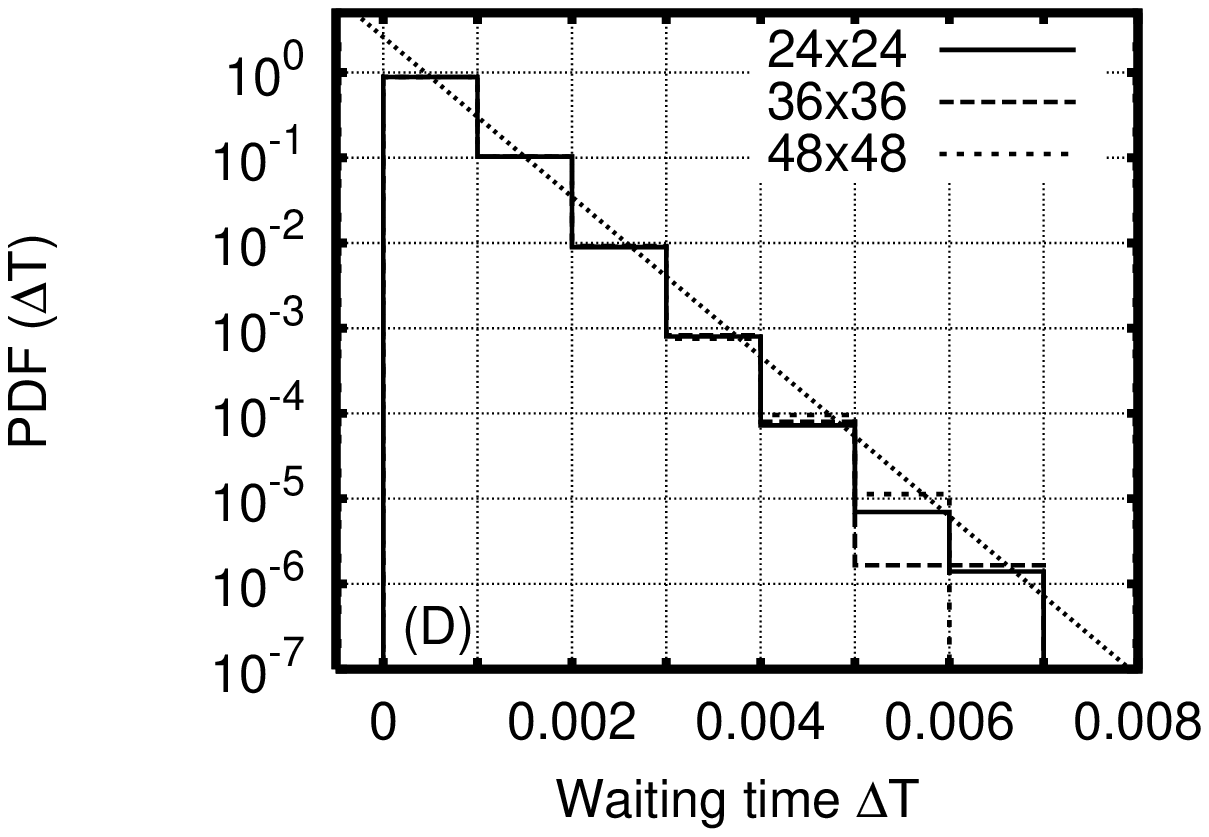}
  \caption{Normalized frequency distributions for the total avalanche energy
    ({\it E}), the peak energy ({\it P}), the avalanche duration ({\it T}) and the waiting
    time~($\Delta T$), for runs of different spatial resolutions
    ($24\times 24$, $36\times 36$, and $48\times 48$) spanning $10^3$
    hyper-diffusion times, containing up to $6\times
    10^5$ avalanches.  Bins of a constant logarithmic width $(\Delta
    \log x) = 0.2$ were used to construct the PDFs.  Power-law indices
    of $\alpha_E=1.407\pm0.02$, $\alpha_P=1.800\pm0.04$, and
    $\alpha_T=2.067\pm0.03$ resulted from fitting data for the
    $48\times 48$ run.  An exponential was fitted to the waiting-time
    distribution, yielding an inverse {\it e}-folding time $\beta=2158\pm4$.}
 \label{fig:pdf}
\end{figure}

In the SOC state,
several characteristics of avalanches have probability
distribution functions (hereafter PDF) that behave as power laws.  This
is the case for
the avalanche energy ($E$), namely the total energy released by the
lattice over the duration of the avalanche; the peak energy release ($P$),
corresponding to
maximum energy released by the avalanche in a single time step; and the
duration ($T$), which is simply the time elapsed from the beginning
of an avalanche to
its end.
Figure~\ref{fig:pdf} (panels A\,--\,C) show
probability distribution functions for $E$, $P$ and $T$, 
constructed from
runs of different resolutions ($24\times 24$, $36\times 36$, and $48\times 48$)
of $10^7$ time steps, spanning 1000 hyper-diffusion times, and
each including up to $6\times 10^5$
avalanches.
Least-squares fit of the form: 
\begin{equation}
  f(x)=f_0\,x^{-\alpha}
\end{equation}
are overplotted on the distributions. The fits used the data from the
large-resolution runs ($48\times 48$).  Because the mesh spacing affects the
statistics for small avalanches, the first few bins are excluded from
these fits.  The power law indices obtained were
$\alpha_E=1.407\pm0.02$, $\alpha_P=1.800\pm0.04$, and
$\alpha_T=2.067\pm0.03$.  These first two values are in good agreement
with those
obtained with the discrete formulation of this avalanche model (see Table~II in
\cite{charbonneau}). The $\alpha_T$ value is a bit higher, but is
in fact more comparable with the values obtained by \cite{lu93} (Table~1).  Figure~\ref{fig:pdf} (panel D) show the frequency 
distribution of the waiting time (WTD; now on semi-log scale).  The
waiting time~($\Delta T$) is the time interval during which the system
is in a stable state between two consecutive avalanches.  The WTD is
well-fitted with exponential of the form:
\begin{equation}
  f(\Delta t)=f_0\,e^{-\beta\Delta t}~,
\end{equation}
here with an inverse $e$-folding
timescale $\beta=2158\pm4$. 
The exponential form of the WTD comes from the statistical
uniformity of the
external forcing, here a stationary random process
\cite[see][]{wheatland}. Other types of forcings can be introduced
in SOC avalanche models for solar flares, producing WTD distributions
in better agreement with observations \cite[see, {\it e.g.},][]{norman},
but for the purposes of the forthcoming validation exercise
we retain the uniform driver of the original Lu and Hamilton model.

To sum up, and notwithstanding some subtleties related
to mesh refinement, 
numerical solutions of the reverse-engineered avalanche
equation lead to avalanching behaviour essentially identical to
that of the original, discrete model of \cite{lu91}.
This ``validation'' of the continuous model may well appear
tautological, but it represents an essential starting point
to our forecasting scheme because recasting the model in 
continuous form is required by the 4D-VAR data assimilation
formalism, the topic to which we now turn.

\section{Data Assimilation (4D-VAR)}
\label{sec:4dvar}

The underlying idea of data assimilation is to use observed data to
introduce corrections to a model --- often a numerical simulation ---
of a physical
phenomenon, typically with the aim of correcting for missing data or to
produce forecasts.  Generally speaking, data-assimilation methods can
be subdivided into three categories: 
successive-correction, sequential (Kalman filter) and variational
methods~\citep{ledimet,daley,kalnay2003}. The successive-correction
method, also known as Cressman method, consists in the correction of a
background field, previously obtained through a previous forecast or a
trivial state due to physical constraints, until it includes the given
observations~\citep{cressman}.  Most sequential methods are based on
the Kalman filter, which uses the model error and the observational
error statistics to find the optimal combination of the model and
observational data~\citep{kantha}.  The variational methods consist in
finding the space-time trajectory of the state variables that will
minimize a cost function measuring the discrepancy between the
forecast and the observations~\citep{talagrand,courtier}.
The 4D-VAR method belongs to this third class and will be the one used
in this work.  Technically speaking, as our model is in
two-dimensional space, our method should be called (2+1)D-VAR.
However, we will continue to refer to it as being 4D-VAR for
generality purpose and because we are following the philosophy behind
the 4D-VAR method.

\subsection{4D-VAR: An Overview}

Four-dimensional variational data assimilation (hereafter 4D-VAR) is
an efficient technique for incorporating observations in numerical
forecasting models~\citep{talagrand,courtier}.  The 4D-VAR method
consists in minimizing a scalar cost function
measuring the deviation between the forecast and the observations.
The physical fields produced by data assimilation must correspond to
the observations, while abiding to the known physical laws and/or
statistical relations characterizing the system being treated~\citep{ledimet}.

\begin{figure}
  \centering
  \includegraphics[scale=0.9]{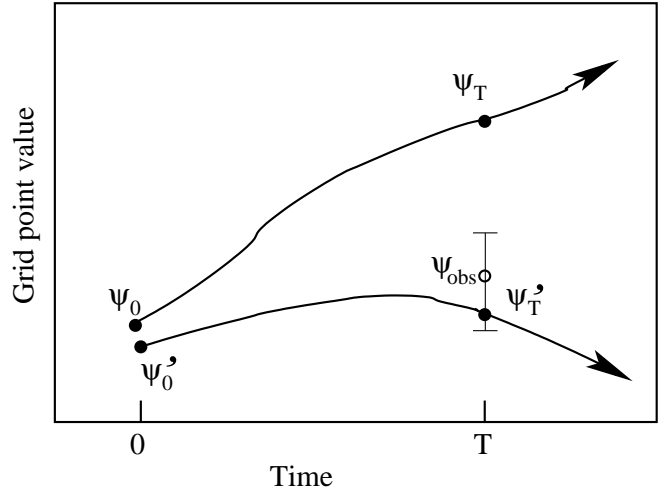}
  \caption{Overview of variational data assimilation.  Point $\psi_0$
    is an estimate of the initial condition at time $0$.  Using this
    initial condition, a forecast $\psi_T$ is obtained at time $T$.
    The 4D-VAR method uses the difference between the forecast
    $\psi_T$ and the observation $\psi_\mathrm{obs}$ to generate a new
    initial condition $\psi_0^\prime$ which will produce a better
    forecast ($\psi_T^\prime$) at time $T$.}
  \label{fig:intro}
\end{figure}

Figure~\ref{fig:intro} shows an overview of the 4D-VAR method as
applied to a classical forecasting problem, namely using the known
state of a variable $\psi_0$ at some initial instant of time, to
forecast its value $\psi_T$ at a later time~$T$. This forecast is to
be compared to an actual observation $\psi_\mathrm{obs}$. Here this
first forecast falls outside of the observation's error bar. The
4D-VAR method uses the difference between the forecast $\psi_T$ and
the observation $\psi_\mathrm{obs}$ to generate a new initial
condition $\psi_0^\prime$ that now produces an improved forecast
$\psi_T^\prime$ at time $T$ which is closer to the
observation~\citep{errico1997}. The procedure can be repeated until
some pre-set goodness-of-fit criterion between $\psi_T^\prime$ and
$\psi_\mathrm{obs}$ is reached. In classical data assimilation
applications (for example numerical weather forecasting), the point
$\psi_0$ would corresponds to a field variable discretized on a single
node $(i,j)$ of a $N\times N$ spatial mesh; 4D-VAR must then solve
concurrently as many variational problems as there are variables
times the number of spatial mesh points on which the numerical
simulation used to advance the field from $0$ to $T$ is performed.
Data assimilation in numerical simulations is a computationally
intensive undertaking!

\subsection{The Cost Function}

Generally, in variational problems, we want to minimize the cost
function $\mathcal{J}$:
\begin{equation}
  \mathcal{J}=\int^{T}_{0}\int_\Omega
  f(\bm{\psi},\mathbf{x},t)\ \d\mathbf{x}\ \d t~,
\end{equation}
where $f(\bm{\psi},\mathbf{x},t)$ is a scalar function, defined over a
domain~$\Omega$ and a time interval~$[0,T]$, of the state
variable~$\bm{\psi}$~\citep{sanders2000}.  More precisely, in data
assimilation, we want to minimize the error between the forecast and
the observations $(\bm{\psi}_T-\bm{\psi}_\mathrm{obs})$ ({\it cf.}~Figure~\ref{fig:intro}):
\begin{equation}
  \mathcal{J} = \frac{1}{2}\int^{T}_{0}\int_\Omega
  (\bm{\psi}_T-\bm{\psi}_\mathrm{obs})^\mathsf{T}\mathbf{W}(\bm{\psi}_T-\bm{\psi}_\mathrm{obs})\
  \d\mathbf{x}\ \d t,
\end{equation}
where $\mathbf{W}$ is a matrix of statistical weights given by the
instrumental errors in the observations and $\mathsf{T}$ indicates the
transpose. The squared residual
$(\bm{\psi}_T-\bm{\psi}_\mathrm{obs})^2$ is used instead of
absolute values, to avoid introducing discontinuities
when the cost function
will be differentiated (next section). The physical equations ({\it i.e.} the continuous avalanche
equation (Equations~(\ref{eq:ava})\,--\,(\ref{eq:ava2}))), which can be
schematically written as:
\begin{equation}
  \label{eq:physeq}
  \mathcal{E}(\bm{\psi},\mathbf{x},t)=0~,
\end{equation}
are acting as constraints during the minimization~\citep{talagrand}.

In this paper, the cost function will use the time series of
released energy as defined via Equation~(\ref{eq:erel}) as the
state variable $\bm{\psi}$.  
Via the use of
Equation~(\ref{eq:elatt}), we get:
\begin{eqnarray}
  \label{eq:er}
  E_r&=&\frac{\d E_l}{\d t}\nonumber\\
  &=&\frac{\d}{\d t}\sum_{i,j}A_{i,j}^2\nonumber\\
  &=&\sum_{i,j}\frac{\d A_{i,j}^2}{\d t}~.
\end{eqnarray}
If we bin the time series of the released energy $E_r$ (Section~\ref{sec:bin}), we have:
\begin{equation}
  \bar{E}_r=\frac{\sum_{\mathrm{bin}}E_r}{\Delta b}~,
\end{equation}
where $\Delta b$ is the number of elements per bin.  The cost function
is then:
\begin{equation}
  \label{eq:cost}
  \mathcal{J} = \frac{1}{2}\int^{T}_{0}
  (\bar{E}_r-\bar{E}_r^\mathrm{obs})^2\ \d t~.
\end{equation}
The identity matrix is used here for the observational error matrix $\mathbf{W}$,
because synthetically-generated observations will be
used~(Section~\ref{sec:result}) in what follows.  Covariance error
matrices are certainly an important and delicate point when
dealing with real data.

\subsection{The Lagrangian Formulation}

We want to minimize the cost function $\mathcal{J}$ given the
constraint $\mathcal{E}(\bm{\psi},\mathbf{x},t)=0$.  Since this is a problem
of minimization with constraints, a Lagrangian formulation is used:
\begin{equation}
  \mathcal{L}(\bm{\psi},\bm{\lambda})=
  \mathcal{J}(\bm{\psi}) + \int^{T}_{0}\int_\Omega
  \bm{\lambda}(\mathbf{x},t)\cdot
  \mathcal{E}(\bm{\psi},\mathbf{x},t)\ \d\mathbf{x}\ \d t~,
\end{equation}
where $\bm{\lambda}(\mathbf{x},t)$ are the Lagrange undetermined multipliers,
also called adjoint variables \break \citep{sanders1999}.  The variational
operator $\delta$ is then applied on the Lagrangian to find its
stationary points:
\begin{eqnarray}
  \delta\mathcal{L} &=&
  \bm{\nabla}_{\bm{\psi}} \mathcal{L}
  \cdot\delta\bm{\psi} +
  \bm{\nabla}_{\bm{\lambda}} \mathcal{L}
  \cdot\delta\bm{\lambda}\nonumber \\
  &=& \diff{\mathcal{L}}{\bm{\psi}}\delta \bm{\psi} +
  \diff{\mathcal{L}}{\bm{\lambda}}\delta \bm{\lambda}~.
\end{eqnarray}
One can use integration by parts to transfer the differential
operators from the state variable $\bm{\psi}$ to the adjoint variable
$\bm{\lambda}$.  For an arbitrary displacement
$(\delta\bm{\psi},\delta\bm{\lambda})$, the minimum is reached only when
$\delta\mathcal{L}=0$~\citep{daley}.  This indicates that the
derivatives of the Lagrangian with respect to each direction must be
zero:
\begin{equation}
  \label{eq:el1}
  \diff{\mathcal{L}}{\bm{\lambda}}=
  \mathcal{E}(\bm{\psi},\mathbf{x},t)=0~,
\end{equation}
and
\begin{equation}
  \label{eq:el2}
  \hspace{0.3in}\diff{\mathcal{L}}{\bm{\psi}} =
  \mbox{Adj}(\bm{\lambda}) + \diff{\mathcal{J}}{\bm{\psi}}=0~,
\end{equation}
where $\mbox{Adj}(\bm{\lambda})$ represents the adjoint
equations~\citep*{schroter}.  

As noted by \cite{ledimet}, this set of equations (Equations~(\ref{eq:el1}) and
(\ref{eq:el2})) are the Euler-Lagrange equations.

For the 2D avalanche model, the direct and adjoint equations are:
\begin{equation}
\begin{array}{ll}\diff{A}{t}\end{array}
=\left\{\begin{array}{ll}-\diff{^2}{x^2}\left(\nu(\nabla^2A)\diff{^2 A}{x^2}\right)
  -\diff{^2}{y^2}\left(\nu(\nabla^2A)\diff{^2 A}{y^2}\right)&{\rm avalanching}\nonumber\\
 F_R(x_0(t),y_0(t))&{\rm stable}\end{array}\right.
\end{equation}
and
\begin{equation}
  \diff{A^*}{\tau}=-\diff{^2}{x^2}\left(\nu(\nabla^2A)\diff{^2 A^*}{x^2}\right)
  -\diff{^2}{y^2}\left(\nu(\nabla^2A)\diff{^2 A^*}{y^2}\right)- \diff{\mathcal{J}}{A}
\end{equation}
respectively, where the generic variable $\bm{\psi}$ has been replaced by
the variable $A$ defined over the lattice, the adjoint variable
$\bm{\lambda}$ has been renamed $A^*$ as it is the adjoint variable
associated with $A$, and $\tau$ is a reverse time ($\tau=T-t$).

Evaluating the term $\diff{\mathcal{J}}{A}$ is particularly delicate
because the cost function $\mathcal{J}$
is not directly given in terms of a spatial quantity related to $A$, but
as a
time series of a nonlinear, non-local function of that quantity:

\begin{eqnarray}
  \label{eq:dJdA}
  \diff{\mathcal{J}}{A}&=&\int^{T}_{0}\frac{1}{2}\diff{(\bar{E}_r-\bar{E}_r^\mathrm{obs})^2}{A}\
  \d t\nonumber\\
  &=&\int^{T}_{0}(\bar{E}_r-\bar{E}_r^\mathrm{obs})\diff{\bar{E}_r}{A}\
  \d t\nonumber\\
  &=&\int^{T}_{0}\frac{(\bar{E}_r-\bar{E}_r^\mathrm{obs})}{\Delta
  b}\diff{\sum_{\mathrm{bin}}E_r}{A}\ \d t~.
\end{eqnarray}
Using the theorem of implicit functions to rewrite the derivative gives:
\begin{equation}
  \label{eq:imp_func}
  \diff{\sum_{\mathrm{bin}}E_r}{A}=-\diff{\sum_{\mathrm{bin}}E_r}{t}
  \left/ \diff{A}{t}\right.
\end{equation}
provided $\diff{A}{t}\ne0$ so the $\diff{\mathcal{J}}{A}$ term is only
to be evaluated when the system is avalanching.  Substituting
Equation~(\ref{eq:imp_func}) in Equation~(\ref{eq:dJdA}):
\begin{eqnarray}
  \diff{\mathcal{J}}{A}&=&-\int^{T}_{0}\left[\frac{(\bar{E}_r-\bar{E}_r^\mathrm{obs})}{\Delta b} 
  \diff{\sum_{\mathrm{bin}}E_r}{t}\left/ \diff{A}{t}\right.\right]\ \d t\nonumber\\
  &=&-\int^{T}_{0}\left[\frac{(\bar{E}_r-\bar{E}_r^\mathrm{obs})}{\Delta b} 
  \left(\sum_{\mathrm{bin}}\diff{E_r}{t}\right)\left/ \diff{A}{t}\right.\right]\ \d t~.
\end{eqnarray}
If we use the definition of $E_r$ (Equation~(\ref{eq:er})), we get:
\begin{equation}
  \diff{\mathcal{J}}{A}=-\int^{T}_{0}\left[\frac{(\bar{E}_r-\bar{E}_r^\mathrm{obs})}{\Delta b} 
  \left(\sum_{\mathrm{bin}}\left(\sum_{i,j} 
\frac{\partial^2A_{i,j}^2}{\partial t^2}\right)\right)
\left/ \diff{A}{t}\right.\right]\ \d t~.
\end{equation}

The initial and boundary conditions for the adjoint equation arise from
the integration by parts, namely the terms evaluated at the limits of
the integrals. The initial conditions are $A^*|_{\tau=0}=0$ and the
boundary conditions are:
\begin{eqnarray}
  A^*(0,y,\tau)=0 &\quad& A^*(L_x,y,\tau)=0 \nonumber\\
  A^*(x,0,\tau)=0 &\quad& A^*(x,L_y,\tau)=0 \nonumber\\
  \left.\diff{A^*}{x}\right|_{x=0}=0&\quad&\phantom{^2}\left.\diff{A^*}{y}\right|_{y=0\phantom{_y}}=0\nonumber\\
  \left.\diff{^2A^*}{x^2}\right|_{x=0}=0&\quad&\left.\diff{^2A^*}{y^2}\right|_{y=0\phantom{_y}}=0\nonumber\\
  \left.\diff{A^*}{x}\right|_{x=L_x}=0&\quad&\phantom{^2}\left.\diff{A^*}{y}\right|_{y=L_y}=0\nonumber\\
  \left.\diff{^2A^*}{x^2}\right|_{x=L_x}=0&\quad&\left.\diff{^2A^*}{y^2}\right|_{y=L_y}=0
\end{eqnarray}

Unfortunately, there is no efficient method to directly solve the
Euler-Lagrange equations (Equations~(\ref{eq:el1}) and (\ref{eq:el2}));
therefore, we must formulate the problem as an unconstrained
problem~\citep{talagrand}. Although we have a random driver in our
model, the model can still be regarded as being deterministic because
the same (arbitrary) sequence of random numbers is being used
throughout the minimization process within 4D-VAR.  Therefore, the
initial conditions, more precisely the ``connected'' portion of 
the system close to the avalanching threshold, will dictate the
evolution of the system with time. This highlights the fact that the
cost function is an implicit function of the initial conditions: it is
by varying the initial conditions that we will find the solution of
the physical equations which minimizes the cost
function~\citep{ehrendorfer}. In the language of optimal control theory, the
initial conditions are the control variable in this
problem~\citep{lions}.

As most minimization algorithms requires the gradient of the function
to be minimized, we need the gradient of the cost function with
respect to the initial conditions.  However, it is not possible to
calculate this gradient analytically as the cost function is an
explicit function of the final conditions ({\it i.e.} forecast).  It turns
out that the more efficient way to calculate the gradient of the 
cost function with respect to the initial conditions is to use the
adjoint equations evaluated at $\tau=T$~\citep{courtier}: 
\begin{equation}
  \label{eq:gradJ}
  \bm{\nabla}\mathcal{J}_{A^0}=A^*(x,y,\tau=T)
\end{equation}
which then requires numerical integration of the adjoint equations
from $\tau=0$ to $\tau=T$, {\it i.e.}, backwards in time from $t=T$
to $t=0$.
This is the reason behind the interest and use of the adjoint
equations, and the need for a continuous form of the avalanche
model. 

\subsection{Minimization Algorithm}

The minimization of the cost function is usually
carried out via a minimization algorithm such as steepest descent, conjugate
gradient, or quasi-Newton methods.
The steepest descent
is a simple method but it converges linearly. In this study, the
conjugate gradient is used because of its quadratic convergence.
The quasi-Newton method also converges
quadratically and is popular among meteorologists.  However, it
requires the computation of the Hessian matrix. Even if an
approximation of the Hessian is normally used, convergence problems
may arise if it becomes nearly singular~\citep{press}.  One can solve
these kind of problem but this leads to an algorithm of a greater
complexity, and it is usually
more computationally intensive.

\begin{figure}
  \centering
  \includegraphics[scale=0.7]{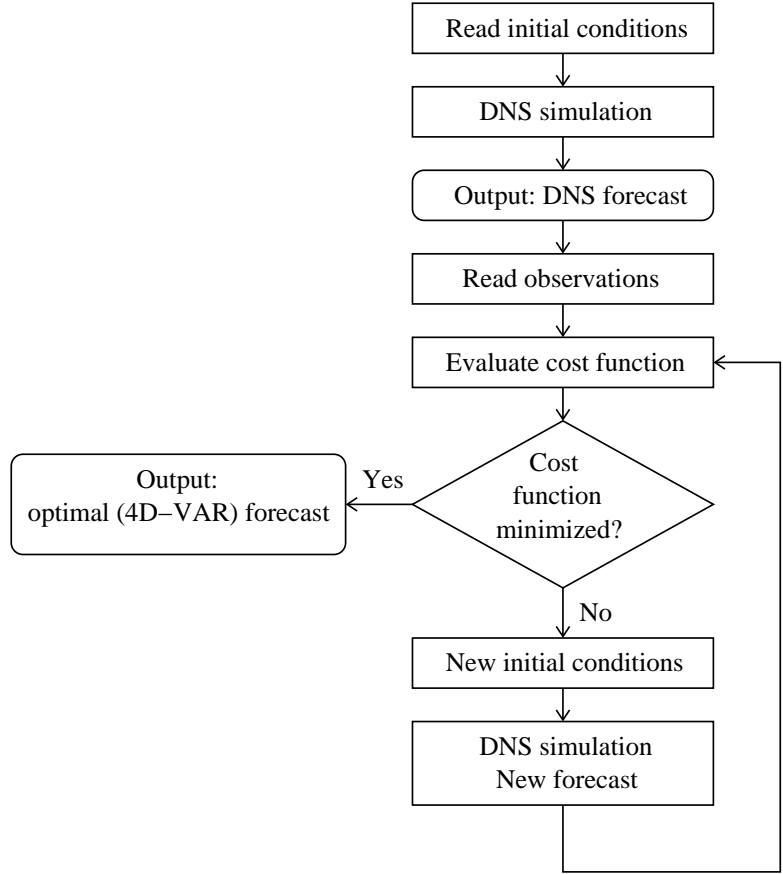}
  \caption{Algorithm of 4D-VAR data assimilation.  From initial
    conditions, a traditional (DNS) forecast is made.  Then the
    observations are read.  The value of the cost function's gradient
    indicates if minimization was achieved. If not, a new set
    of initial conditions are produced by backwards integration
    of the adjoint equations, a new forecast
    is produced, and the
    verification is repeated.  This procedure is iterated until
    minimization of the cost function is achieved. The output is the
    optimal 4D-VAR forecast and associated initial conditions.}
  \label{fig:algo}
\end{figure}

The algorithm implementation of 4D-VAR data assimilation runs as
shown on Figure~\ref{fig:algo}.
Starting from initial conditions obtained by current
experimental observations or a previous numerical simulation, a direct
simulation generates a traditional (DNS)
forecast.  After reading the observations taken
at the end of the forecast period, the cost function is evaluated,
followed by the evaluation of its gradient.  If the gradient is
smaller than a chosen tolerance which accounts for error due to
numerical precision, then the minimum of the cost function has been
reached and the optimal 4D-VAR forecast is obtained.  If
the cost function is not minimized, we must iterate.  A new set of
initial conditions are generated and used as the starting
point of a new forecast. The cost function and its gradient are
reevaluated and checked again against the termination criterion. 
This procedure is repeated until the latter is met.
The iteration loop in Figure~\ref{fig:algo} takes place within the
conjugate gradient minimization algorithm.  It is also the conjugate
gradient which modifies the initial conditions $A^0_{i,j}$ at each
iteration:
\begin{equation}
  [A^0_{i,j}]^{k+1}=[A^0_{i,j}]^{k}+\alpha^kp^k_{i,j}~,
\end{equation}
where, for the $k^\mathrm{th}$ iteration of the conjugate gradient, $p^k_{i,j}$
is the conjugate-direction vector multiplied by an amplitude $\alpha^k$.
These conjugate direction vectors are obtained with the use of the gradient:
\begin{equation}
  p^{k+1}_{i,j}=\bm{\nabla}\mathcal{J}^{k+1}_{i,j}+\gamma^kp^{k}_{i,j}
\end{equation}
where
\begin{equation}
  \gamma^k=\frac{(\bm{\nabla}\mathcal{J}^{k+1}_{i,j}-\bm{\nabla}\mathcal{J}^{k}_{i,j})\cdot\bm{\nabla}\mathcal{J}^{k+1}_{i,j}}{\bm{\nabla}\mathcal{J}^{k}_{i,j}\cdot \bm{\nabla}\mathcal{J}^{k}_{i,j}}~,
\end{equation}
and the gradients are evaluated using Equation~(\ref{eq:gradJ}).
The amplitude $\alpha^k$is calculated with:
\begin{equation}
  \alpha^k=b-\frac{1}{2}\frac{(b-a)^2[\mathcal{J}(b)-\mathcal{J}(c)]-(b-c)^2[\mathcal{J}(b)-\mathcal{J}(a)]}{(b-a)[\mathcal{J}(b)-\mathcal{J}(c)]-(b-c)[\mathcal{J}(b)-\mathcal{J}(a)]}
\end{equation}
where $a$, $b$, and $c$ are modified initial conditions that where
obtained through a line minimization method known as inverse parabolic interpolation, which iteratively
finds a triplet of points such that the minimum of a
parabola passing through these three points will be as close as
possible to the minimum of the function in that given interval~\citep{press}.

\subsection{Beyond Classical 4D-VAR}

There are many ways in which our use of 4D-VAR goes beyond the
``classical'' formulation of 4D-VAR, as given on Figure~\ref{fig:intro}.
We are using 4D-VAR to assimilate data into a cellular automaton made
continuous only for the purpose of 
writing adjoint equations. Grid and time steps are fractions of the
characteristic scales of the hyper-diffusive processes involved
and are not infinitesimals.
However, as pointed out by~\cite*{isliker}, it is still possible to compute
derivatives and thus operators. To the best of our knowledge,
at the present time the only other area in which data assimilation
techniques are being developed 
in conjunction with a cellular automaton is in seismic
data assimilation of a stochastic random fault model~\citep{rundle,gonzalez}.

We are also using a time series of a global, model-produced variable
to define the error, as opposed to the spatial state of the system
measured at some time interval beyond the initial condition
in which data is being assimilated.
Assimilating time series of a global variable instead of (or together
with) spatial states of a system at non-zero times $t$ is truly
compatible with the 4D-VAR approach, as opposed to 3D-VAR for instance.
Models in
environmental sciences have been, and will increasingly be, assimilating
time-series of data (see for instance~\cite*{carton} in Oceanology,
\cite*{bertino} in Estuary modeling, or \cite{eymin} in
Geomagnetism).

Finally, our model system includes an essential stochastic component,
namely the driving.
Ocean and atmosphere general circulation or solid-Earth dynamics are all
random driven (noisy) systems. Data
assimilation, including 4D-VAR, in principle can be applied to any such system
\cite[{\it e.g.},][]{nichols,moore}.  Random driving,
inaccessible to observations, cannot be assimilated, but is not
seen as a systematic bias in the model \cite[see, {\it e.g.},][]{anderson}.

\section{Validation Experiments}
\label{sec:result}

\subsection{Experimental Design}

\subsubsection{Synthetic Data}

The validation experiments use synthetic observations generated from the same
SOC ava\-lanche model used to carry out data assimilation (certainly
an optimal situation from the data assimilation point of view).
Of course, entirely independent realizations of the stochastic driving
are used to generate the synthetic observations, and these
realizations are {\it not} made available to 4D-VAR, as our challenge
is precisely to see whether 4D-VAR can still adequately assimilate
the observations without ``knowing'' about the stochastic driving.

The synthetic energy-release time series cover one hyper-diffusion time,
and were produced on a $48\times 48$ grid, with time step $\Delta t/\tau_\nu= 5.5\times 10^{-11}$,
and forcing parameters as given in Section~2.2.2.

\subsubsection{Thresholding and Binning}
\label{sec:bin}

\begin{figure}
  \centering
  \fbox{\includegraphics[scale=0.16]{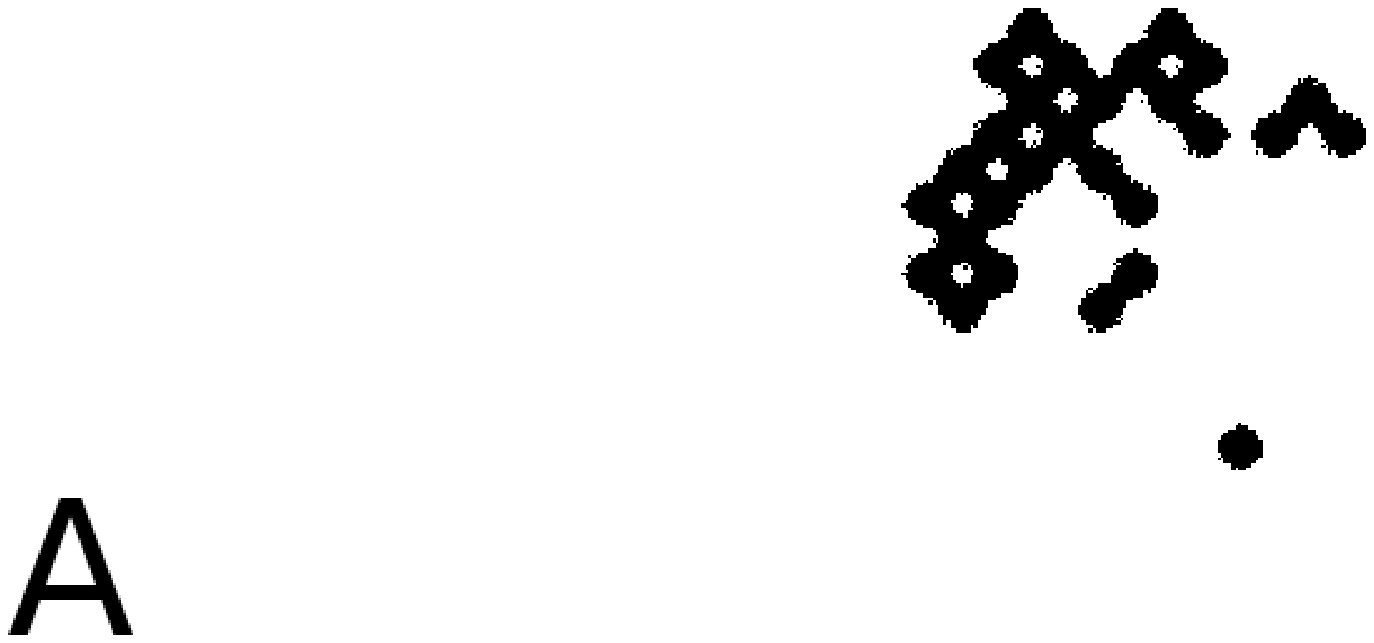}}
  \fbox{\includegraphics[scale=0.16]{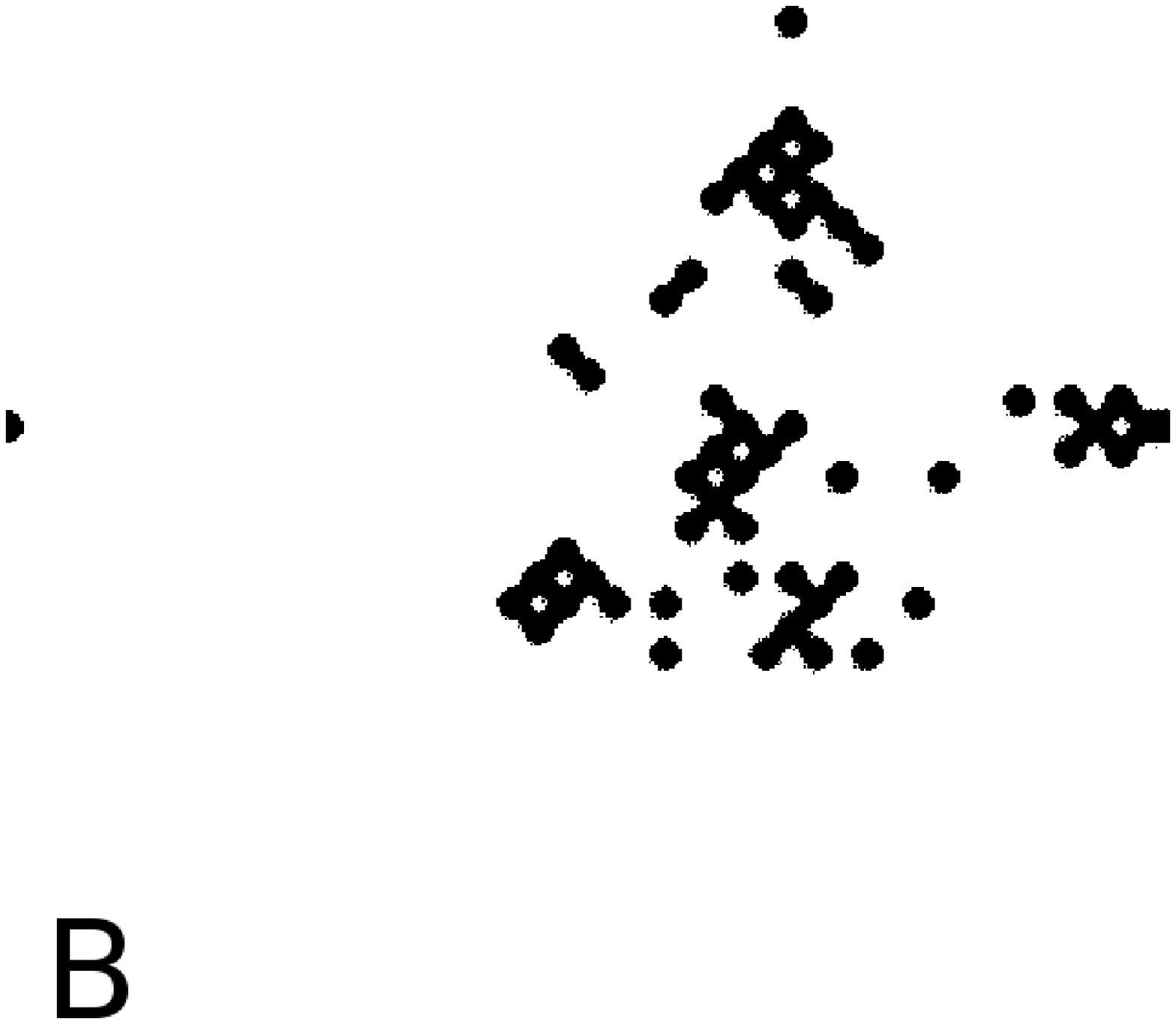}}
  \fbox{\includegraphics[scale=0.16]{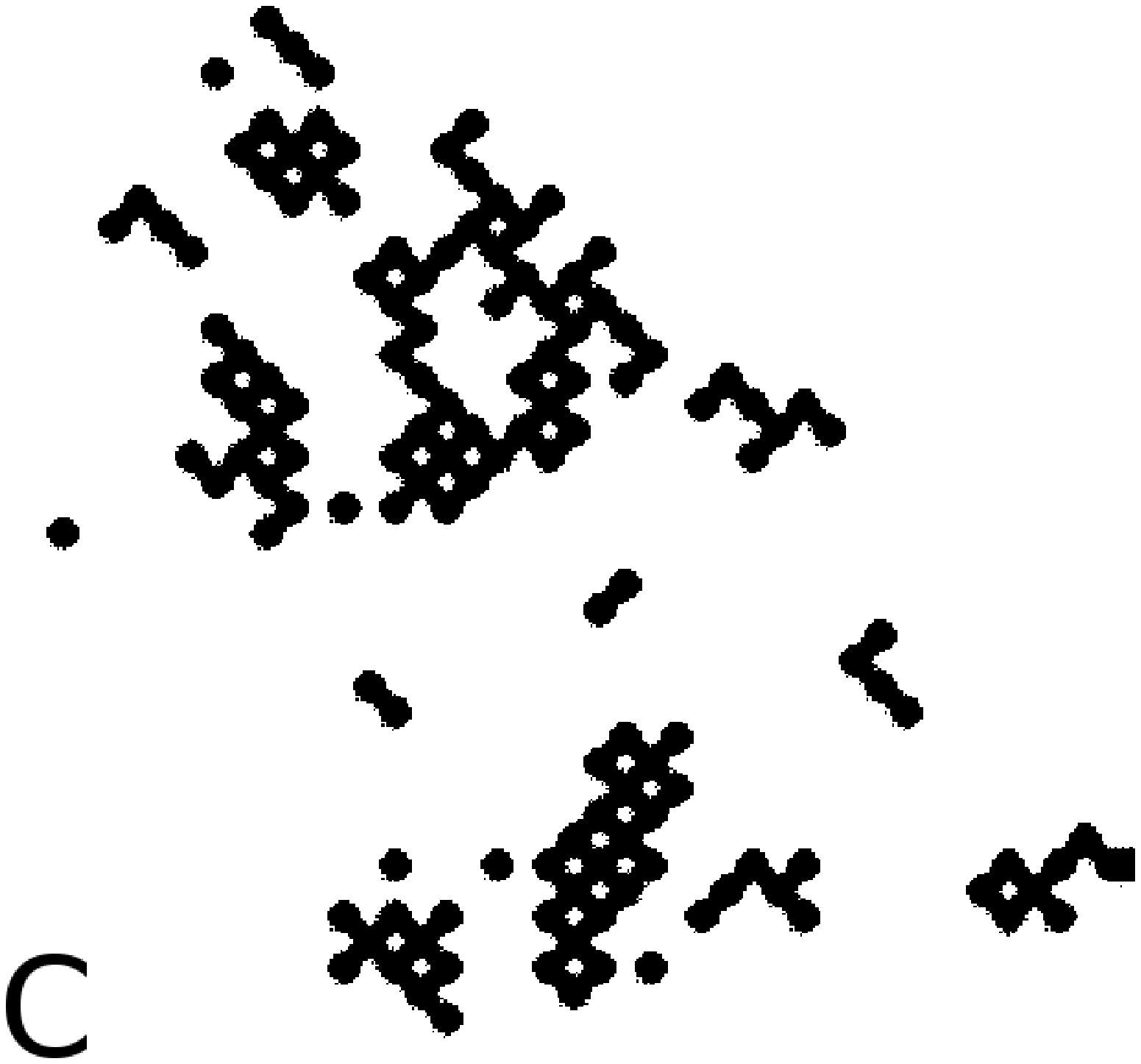}}
  \fbox{\includegraphics[scale=0.16]{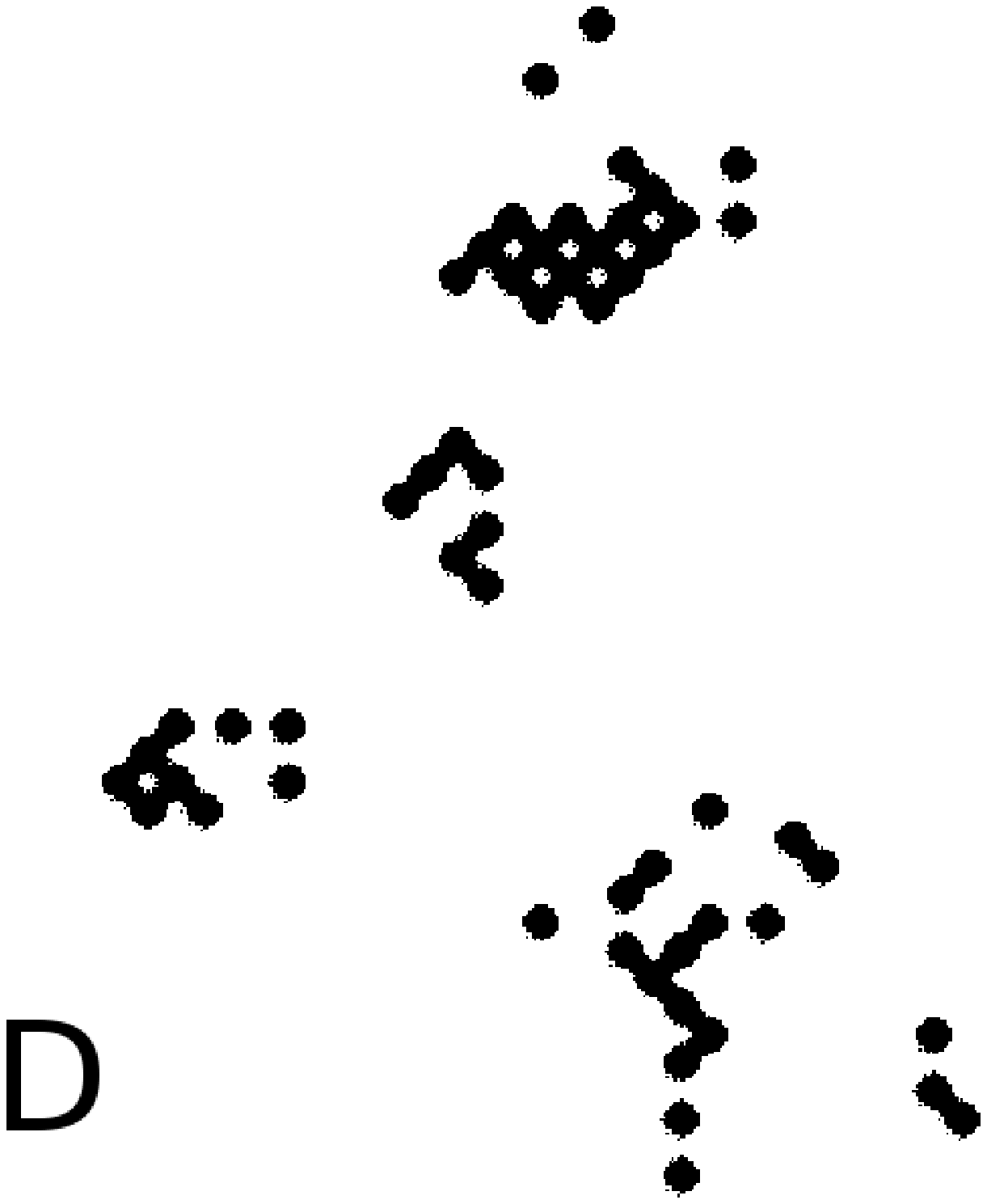}}
  \includegraphics[scale=0.96]{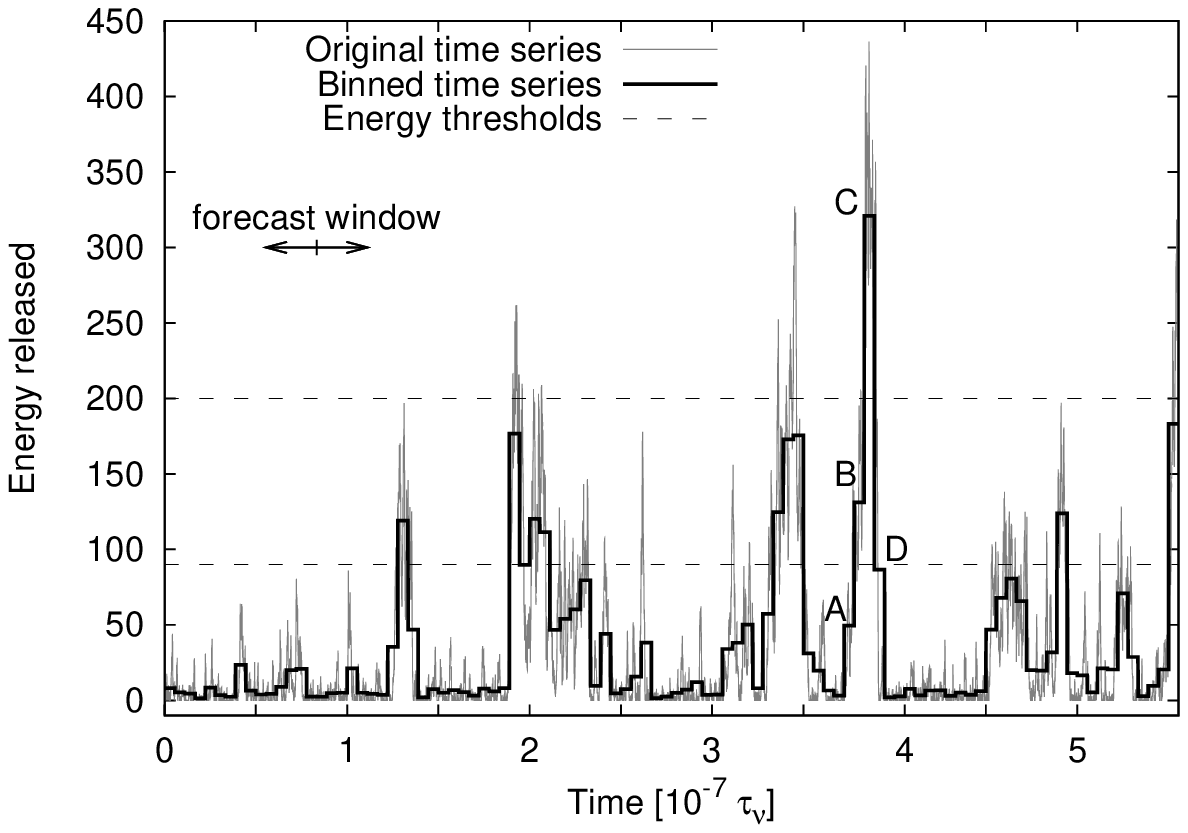}
  \caption{A small segment of a run of the avalanche model.  Several
    aspects of the experimental design are displayed in 
    the bottom panel.  The gray line is an original time
    series of released energy generated by the avalanche model
    (Equation~(\ref{eq:ava})). The thick solid line is
    the same series which has been binned by using bins of width of
    $\Delta b=100$ time steps.  Only the avalanches above the energy threshold (horizontal
    dashed lines) will be considered. In the top left corner of the
    main panel, a typical
    forecast window is shown.  It determines the maximum time interval
    between an observed and forecasted avalanche in order to have a match.
    The top four panels show the avalanching regions evolving
    in space and time. From the initial perturbation at a random
    point, the avalanching region increases from a small region (A) to
    a large region occupying an important portion of the domain where
    the hyper-diffusion has reached the sides of the domain (B and C).
    The avalanche is typically fragmented in its decay phase (D).}
  \label{fig:Erel-avg}
\end{figure}

Figure~\ref{fig:Erel-avg} illustrates a segment of representative simulation run. The
gray line in the main panel of Figure~\ref{fig:Erel-avg} is a typical
time series for the energy released by the avalanches. Time is measured in
units of the magnetic hyper-diffusion time scale, and energy in
arbitrary units. This arbitrary energy scale can be mapped to the
standard flare classification by dividing the peak energy that covers
three decades (panel B of Figure~\ref{fig:pdf}), in four ranges equal in
logarithmic size. Thus, the following association can be made: $P<8$
are B-class flares, $8<P<40$ are C-class flares, $40<P<200$ are
M-class flares, and $P>200$ are X-class flares.

Data assimilation is carried out on a binned version of the energy
time series, shown as a thick solid line in Figure~\ref{fig:Erel-avg}.
A bin width of $\Delta b=100$ ({\it i.e.} 100~time steps) was chosen,
as it is
large enough to remove the small structures but small enough to keep
the general features of the avalanche.  This binning facilitates the
minimization of the cost function by eliminating the fine structure
details, which are unnecessary as we are mostly interested in
forecasting the flaring time, peak flux, and total released energy.

Two parameters are used to build statistics.  First, there is an
energy threshold (horizontal dashed lines in Figure~\ref{fig:Erel-avg}).
Only avalanches with an energy above the threshold are considered in
the statistics.  The signals below the threshold are treated as being
noise or low energy avalanches that are unimportant.  Here the two
thresholds that will be used in this paper are shown: a threshold at
90 and a larger one at a value of 200 energy units. The lower
threshold retains the model's equivalent of upper half of the medium
size (M-class) and large size (X-class) flares while the larger one
will only take the high energy X-flares into account. These classes of
flares are the ones having the most important effects on Earth.  The
second parameter is the forecast window. A forecast window of a range
of $\delta t=0.55\times 10^{-7}$ is depicted in the upper left corner
of the main panel of Figure~\ref{fig:Erel-avg}.  The forecast window is the maximum time
interval between an observed peak and a forecasted peak to have a
match.  In the case of a forecast window of $\delta t=0.55\times 10^{-7}$, the predicted
peak can either be $\delta t=0.275\times 10^{-7}$ before or after the observed peak.  Hence,
a smaller forecast window implies a more precise forecast.

The top four panels of Figure~\ref{fig:Erel-avg} show the evolution of
the avalanching regions at different stages throughout the spatio-temporal
evolution of a large avalanche.
Starting from an initial perturbation at a given
random point, the ``preflare'' stage begins with an avalanching region
occupying a small region (panel A, Figure~\ref{fig:Erel-avg}) of the
domain that will gradually increase (panel B, Figure~\ref{fig:Erel-avg})
to reach a large size, occupying
an important portion of the domain, at the ``impulsive stage''
at the end of which the
peak is reached (panel C, Figure~\ref{fig:Erel-avg}).  At this
point, the fragmented hyper-diffusion front
has reached the sides of the domain.  In
the ``decay'' stage, the avalanching regions start to decrease in size (panel D,
Figure~\ref{fig:Erel-avg}).  Eventually, the system will return to a
stable state, and driving resumes until the triggering of
the next flare/avalanche.

\subsubsection{Maintaining the SOC State}

To prevent the system from leaving the SOC state due to the 4D-VAR
correction made to the initial conditions, the lattice energy of the
corrected initial conditions is compared to the average lattice
energy, $\bar{E_l}$, of the SOC state.  If the lattice energy of the
corrected initial conditions is not found within the variance
$\sigma$, the 4D-VAR correction is adjusted by a factor $\epsilon$
defined as:
\begin{equation}
  \begin{array}{l}\epsilon\end{array}
  =\left\{
    \begin{array}{ll}
      -\frac{|E_l-\bar{E_l}|}{\sigma}, &\quad\mathrm{if}\ E_l>\bar{E_l}+\sigma\\
      1, &\quad\mathrm{if}\ \bar{E_l}-\sigma<E_l<\bar{E_l}+\sigma\\
      \frac{|E_l-\bar{E_l}|}{\sigma}, &\quad\mathrm{if}\ E_l<\bar{E_l}-\sigma
    \end{array}
  \right.
\end{equation}
which brings the initial conditions' lattice energy within the variance.

\subsubsection{Random Number Sequences}

For the minimization procedure within the 4D-VAR
algorithm, the same seed is used to initialize the random number
generator, so that during each iteration within 4D-VAR
({\it cf.}~Figure~\ref{fig:algo}),
the same sequence of perturbations is added in the same order to
the same mesh points.

\subsubsection{Running 4D-VAR}

\begin{figure}
  \centering
  \includegraphics[scale=0.94]{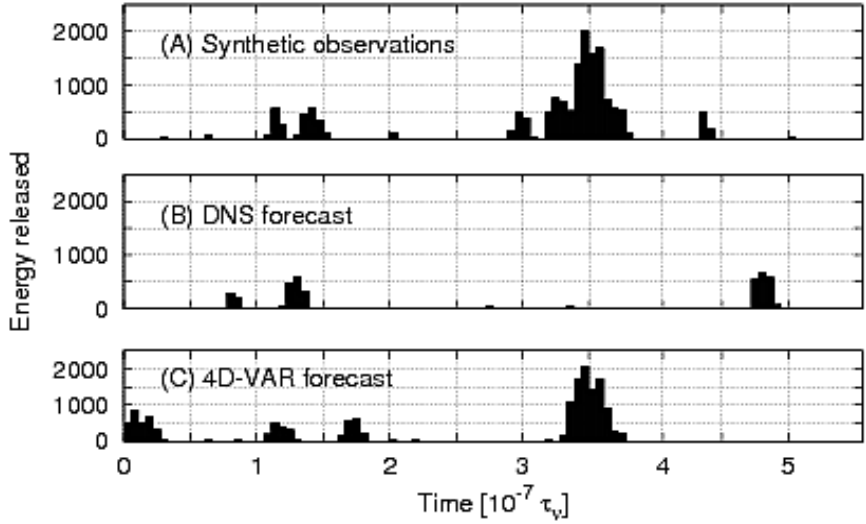}
  \caption{A sample run of 4D-VAR data assimilation.
    The
    synthetic observations being assimilated are in panel A.
    Even using the same initial condition $A(x,y,t=0)$,
    a single DNS ``forecast'' (panel B) often
    produces poor results, a direct consequence
    of the stochastic nature of the driving
    process. Retaining the same driving but allowing 4D-VAR to alter
    the initial condition (panel C) results in
    a much better representation of the observations.}
  \label{fig:result}
\end{figure}

With a realization of stochastic driving different from the one that
has been used to
produce the synthetic observations (panel A, Figure~\ref{fig:result}),
a DNS ``forecast'' is produced (panel B, Figure~\ref{fig:result}). The same
initial conditions were used in both cases, but the different realizations
of the driving have led, perhaps not surprisingly, to very different
time series of energy release.  Although the DNS run has
reproduced the small avalanche at $t=1.1\times 10^{-7}$, it missed the large one at
$t=3.5\times 10^{-7}$.  The 4D-VAR run (panel C, Figure~\ref{fig:result}),
using the same driving realization as the DNS forecast but with
corrected initial conditions, has correctly reproduced the large avalanche.
Figure~\ref{fig:result} is a typical case when the 4D-VAR method
performs well. There are cases where the DNS is already quite good, and
4D-VAR cannot produce significant improvement; this is in fact expected, and moreover is the
reason why true forecasting may be possible despite the stochastic
nature of the forcing (more on this in the concluding section).
In this representative sample run, minimization of the cost function
by 4D-VAR was achieved in a mere five iterations of the conjugate gradient.  This 4D-VAR procedure can thus be summarized as follows:
\begin{itemize}
\item Production of synthetic observations with a given sequence of random numbers
\item A DNS run using a different sequence of random numbers is performed
\item A 4D-VAR run with the same sequence of random numbers as used
  for the DNS run is performed
\item As the 4D-VAR method benefits from corrected initial conditions, it is more successful than the DNS run in reproducing the observations
\end{itemize}

\subsection{Performance}

\subsubsection{Performance Measurements}

In anticipation of true forecasts, it is instructive to analyze
the performance
of the 4D-VAR runs in terms of matches,
misses, and false alarms.  Only the avalanches with a peak above the energy
threshold are considered.  With the forecast window centered at the
peak of each observed avalanche, the forecast is examined to see if
one of its avalanches takes place inside the window.  If this is the case,
we have a match. If we use an energy threshold of 200 and a forecast
window of $\delta t=0.55\times 10^{-7}$, Figure~\ref{fig:result} has two matches: the large
avalanche at $t=3.4\times 10^{-7}$ and a smaller one at $t=1.1\times 10^{-7}$.  The avalanche
at $t=1.1\times 10^{-7}$ is a possible match for either one of the avalanches at $t=1.1\times 10^{-7}$ and
$t=1.38\times 10^{-7}$.  Such situations are treated as a single match.  Hence, a
match can be considered as an event happening in a time
interval, determined by the forecast window, regardless if there is a
single avalanche or multiple consecutive avalanches.  A miss happens when the
observed avalanche does not have a counterpart in the 4D-VAR run.
Figure~\ref{fig:result} (panel C) has two misses: at $t=2.9\times 10^{-7}$ and $t=4.3\times 10^{-7}$.
False alarms are avalanches appearing in the 4D-VAR run
which do not have counterparts
in the observations.  Figure~\ref{fig:result} (panel C) has one false
alarm at $t=1.7\times 10^{-7}$.  The avalanche at $t=0.1\times 10^{-7}$ is not considered
as being a false alarm as it is an artifact of the 4D-VAR method. The
correction to the initial conditions has left the system in an
unstable state so the forecast time series began with an avalanche.
Finally, the two small avalanches at $t=0.6\times 10^{-7}$ and $t=2\times 10^{-7}$ in the
observations are not included in the analysis as they fall below the
energy threshold.

The final value of the cost function is not necessarily
an optimal measure of a successful run, as it simply
measures the mean quadratic difference between the observation
and model time series over the whole duration of the assimilation
interval.
In the flare-forecasting context, one is primarily interested in
predicting the timing of discrete events, namely the largest
flares/avalanches, and ideally also a measure of their peak flux and/or
total released energy.
Consequently, we define the following
quality factor ($Q$)
to assess numerically whether a
given run was successful or not in ``catching'' avalanches in the observations:
\begin{equation}
  \label{eq:sigma}
  Q=\alpha\sum_{\mathrm{match}}\left|\frac{E_o}{E_o-E_f}\right|
  -\beta\sum_{\mathrm{miss}}\left(\frac{E_o}{E_{tot}}\right)
  -\gamma\sum_{\mathrm{false\ alarm}}\left(\frac{E_f}{E_{tot}}\right)~.
\end{equation}
In Equation~(\ref{eq:sigma}), the first term is the sum, for all pair of
matching flares, of the inverse of the error difference between the
total energy of the observed ($E_o$) and modeled ($E_f$) avalanches
multiplied by a factor $\alpha$. The two next terms are penalties due
to the misses and false alarms. The penalty are defined as the energy
of the missed ($E_o$) and false alarm ($E_f$) avalanches normalized by the
total energy released by all the observed avalanches ($E_{tot}$). These
three individual contributions to $Q$ are then each assigned a distinct
weighting factor, chosen here as $\alpha=4$, $\beta=2$, and
$\gamma=1$, so that the largest weight is for the matches. Note also that under
these definitions, missing a large avalanche incurs a larger penalty
than missing a small one. The misses have a larger
weight than the false alarms because a miss leaves us unprepared to
handle the consequence of the flare.  On the other hand, in a false
alarm we may incur additional costs even though no flare is triggered.

\subsubsection{Performance Statistics}

Statistics of hits, misses, and false alarms has been gathered for 100
4D-VAR runs. These runs have been realized with combinations of ten
sets of observations and ten sets of distinct random number sequences
each setting a distinct realization of stochastic driving within the
SOC model. For all runs included in the statistical analysis to
follow, an energy threshold of 90 and a forecast window of $\delta t=0.55\times 10^{-7}$ were
used.

\begin{table}
  \centering
  \begin{tabular}{c|c|c|c|c|c}\hline
    \multirow{3}{0.62in}{\centering Number of Consecutive Avalanches} 
    & \multirow{3}{0.5in}{\centering Number of Matches} 
    & \multirow{3}{0.4in}{\centering Number of Misses} 
    & \multicolumn{2}{|c|}{Number of False Alarms}
    & \multirow{3}{0.86in}{\centering Number of Avalanches Below Threshold}\\
    &&& \multicolumn{2}{|c|}{since Last Match}&\\\cline{4-5}
    &&&\centering False Alarms &\centering Number of Runs&\\\hline\hline
    \multirow{5}{*}{1}&\multirow{5}{*}{}&\multirow{5}{*}{}&0&44 [26]&\multirow{5}{*}{}\\
    &&&1&12 [9]&\\
    &65 [41]&35 [59]&2&8 [3]&0 [0]\\
    &&&3&0 [2]&\\
    &&&4&1 [1]&\\\hline
    \multirow{3}{*}{2}&\multirow{3}{*}{27 [22]}&\multirow{3}{*}{28
      [14]}&0&22 [15]&\multirow{3}{*}{10 [5]}\\
    &&&1&4 [3]&\\
    &&&2&1 [4]&\\\hline
    \multirow{3}{*}{3}&\multirow{3}{*}{11 [13]}&\multirow{3}{*}{7
      [6]}&0&10 [8]&\multirow{3}{*}{9 [3]}\\
    &&&1&1 [4]&\\
    &&&2&0 [1]&\\\hline
    4&8 [1]&1 [5]&0&8 [1]&2 [7]\\\hline
    5&2 [0]&0 [0]&0&2 [0]&6 [1]\\\hline
    6&2 [0]&0 [0]&0&2 [0]&0 [0]\\\hline
    7&0 [0]&0 [0]&0&0 [0]&2 [0]\\\hline
  \end{tabular}
  \caption{Performance of the system in matching $n$ consecutive
    avalanches before the first miss. The results from the regular DNS
    runs (numbers in square brackets) were added  for comparison
    purpose.  The second and third columns lists the number of runs that
    had a match or a miss for the $n^\mathrm{th}$ avalanche,
    respectively. Fourth and fifth columns displays the number of
    false alarms between the $n^\mathrm{th}$ and $(n-1)^\mathrm{th}$
    avalanches and the corresponding numbers of runs which had these false
    alarms.  The sixth column keeps track of the runs that have neither a
    match nor a miss because there are no longer any avalanches above
    the threshold.}
  \label{tab:consecmatch}
\end{table}

The idea here is to investigate the performance of the system 
{\it over the assimilation interval} (we are still not forecasting at
this stage), in terms of the number of consecutive matches that could
be obtained before the first miss.  The results are tabulated in
Table~\ref{tab:consecmatch} for each $n^\mathrm{th}$ avalanche (first
column).  The second and third columns lists the number of runs that
had a match or a miss for the $n^\mathrm{th}$ avalanche, respectively. The
fourth and fifth columns displays the number of false alarms between
the $n^\mathrm{th}$ and $(n-1)^\mathrm{th}$ avalanches and the number
of runs which had these false alarms.  The last column accounts for
the fact that the runs do not have the same number of avalanches above
the threshold.  The results of applying the same analysis to regular
DNS runs are added for
comparison purpose (numbers in square brackets).

Examining the first row, we see that, of the 100 runs, 65 4D-VAR runs
succeeded in reproducing the first avalanche (above threshold) while 35
runs missed it. Of the 65 runs with the match, 44 had no false alarms
between the beginning of the run to the first avalanche while 21 of them
had between one and four false alarms in this time interval.  The DNS runs
were less successful in reproducing the first avalanche as 41 of them were
able to do so. The proportion of false alarms before the first avalanche
is similar for both the 4D-VAR and the DNS runs.  Every
run, 4D-VAR and DNS, had at least one avalanche above the threshold.  This
first result is quite interesting: 44\% of the 4D-VAR runs
successfully reproduced the first observed avalanche without making a false
alarm; this is a 60\% increase in performance compared to the DNS
runs.  If we move on to the second row, of the 65 4D-VAR runs which
reproduced the first avalanche, 27 of them were also able to reproduce the
second avalanche.  The remaining 38 runs either missed the second avalanche
(28 runs) or did not had a second avalanche higher than the threshold (10
runs).  The number of false alarms between the first and second avalanches
follows the same trends as the ones before the first avalanche. The
difference between the 4D-VAR and DNS runs is less pronounced as 22
DNS runs matched the second avalanche.  This implies that the 4D-VAR
method is very good at reproducing the first avalanches but afterward the
performance degrades to become equivalent to the DNS method.
However, the
4D-VAR method did reproduce runs of five and six consecutive avalanche when
the DNS runs no longer produced avalanches above the energy threshold.
Finally, the number of consecutive matches continues to decrease until
either an avalanche is missed or all observed avalanches are reproduced.

The reason behind this somewhat sudden decrease in performance comes
from the fact that avalanches, especially large ones, have a deep impact
on the energy distribution on the lattice. Thus, even with synthetic
data produced by the same model used for data assimilation, it become
more difficult to reproduce the next avalanche, which explains the
constant decrease in the number of matches. This is a direct
consequence of the stochastic nature of the driving process, and
already heralds the finite forecasting window that can be expected
when operating in true-forecasting mode.  Nevertheless, one tenth of
the total 100 runs could still reproduce the first three avalanches. Only
two runs reproduced the first six avalanches above threshold (although 61
runs did not have a sixth avalanche to reproduce).  However, at this
point, the size of the time interval for the run ($t=5.5\times 10^{-7}$) is felt as
nine runs that match the first two runs did not have a third avalanche
above the threshold.  The runs used here have an average of four avalanches above the
energy threshold.

\subsubsection{Code Performance}

At a spatial resolution of $48 \times 48$, 
it takes about five minutes of wallclock time to complete a data
assimilation run over 10\,000 time steps
on an Intel Itanium~2 processor.
Note that no particular efforts were made towards code optimization.
Although the writing of the adjoint equations
in the 4D-VAR implementation can be difficult, the resulting data
assimilation scheme
is quite fast. The hope is that this performance
will not degrade too much once real data, including observational
errors and covariance matrices, will be used for true flare forecasting.

\section{Conclusion: Towards Forecasting}
\label{sec:conclusion}

What have we actually achieved with this whole data assimilation procedure?
Let us go back to the idea of predicting flare occurrence and energy release
via a direct numerical simulation based on a SOC
avalanche model. To run such a model forward in time,
two things
must be specified: {\it i}) the current state of the lattice at time
$t_0$,
to be used as initial condition for the forecast, and {\it ii})
the spatio-temporal sequence of perturbations throughout
the forecasting interval. If the latter are truly stochastic in
nature, they remain completely unknown at $t_0$. In the
context of Parker's nano-flare hypothesis, which provides the
physical underpinning of avalanche models in the present situation,
these perturbations amount to small kinks between adjacent
magnetic fieldlines, building up in response to slow
forcing of the structure's photospheric magnetic footpoints.
Not only do these kinks develop in response to stochastic
forcing, but they also occur on spatial scales
inaccessible to direct observation. This means that
flare forecasting using an avalanche model will always
retain a stochastic component.

What we have shown in this
paper is that past avalanching behaviour can be reproduced
reasonably well using data assimilation, {\it even without
detailed knowledge of the stochastic forcing}. At the end of the
assimilation interval, the lattice is in a state that
is compatible with (and determined by) past flaring behaviour.
This then represents the optimal initial condition from which
to carry out a DNS forecast. This, of course, does not guarantee that
any given DNS forecast will be accurate, but that an ensemble
of DNS forecasts will show avalanching patterns that reflect,
at least in part, the state of the lattice at $t_0$.
In particular, if this initial condition is
characterized by a large, connected portion of the lattice
close to the stability threshold, then one would expect
that a large avalanche is
likely in the near future, irrespective of the spatio-temporal
details of the forcing. It should then be possible to forecast
with some accuracy the largest upcoming avalanches using
statistical ensembles of DNS runs. Small avalanches, on the
other hand, will depend more sensitively on details of the
(stochastic) forcing. In such cases, even ensembles of DNS runs are less
likely to produce useful forecasts. In the space-weather context,
this is not too problematic, since it is precisely the largest flares/avalanches
for which one is seeking accurate forecasts.
These expectation are examined in detail in the
following paper in this series~(Bélanger, Vincent, and Charbonneau, in
preparation).

SOC avalanche models are certainly not the only modeling
framework for solar flares
within which data assimilation can be carried out.
A good case in point is the CISM project \citep{wiltberger,siscoe},
an ambitious sun-to-ionosphere data assimilation 
framework based on a suite of coupled 3D MHD models.
The attractive feature of SOC models --- arguably their single
most attractive feature --- is that,
by all appearance, they correctly capture the
global statistical behaviour of energy release by solar flares,
including in particular its power-law form and
associated exponent. This makes such models ideal
candidates for data-assimilation-based forecasting,
despite their extreme physical simplicity and
the inevitable stochastic effects associated
with the driving mechanism responsible for energy
injection into the overlying coronal magnetic structures.

We note, in closing, that within Parker's physical picture
of coronal structures being forced by photospheric fluid motions,
such stochastic effects would also need to be incorporated
into any full-scale MHD models of coronal structures to be used
for flare forecasting. Data assimilation could help here as
well \cite[see, {\it e.g.},][]{schrijver}, but spatially- and/or
temporally-unresolved fluid motions would again introduce
a form of stochastic ``noise'' in the MHD simulations,
with inevitable degradation of
forecasting performance even if such models would be based
on true physical equations rather than some largely {\it ad hoc}
cellular automaton. The latter, however, is such a simpler
model to simulate that it becomes possible in practice
to carry out ensemble DNS forecasting in reasonable wallclock time
even on mid-range computational platforms. This is an essential
requirement of operational forecasting.

\acknowledgements
The authors would like to acknowledge financial support of the \break
FQRNT-NATEQ (Projet de recherche en équipe: Avalanches et prédiction
des éruptions solaires) and NSERC Discovery Grant grants to A.V.
and P.C. The
authors are also grateful to RQCHP for providing computational
support.

\bibliographystyle{aa}
\bibliography{article_avalanche}

\end{article}
\end{document}